\documentclass[12pt]{iopart}

\usepackage{graphicx}
\usepackage{epsfig}

\begin{document}

\title{Morse-like squeezed coherent states and some of their properties}
\author{O de los Santos-S\'anchez$^{(1)}$ and J R\'ecamier$^{(2)}$} 

\address{$^{(1)}$ Instituto de F\'{\i}sica, Universidad Nacional Aut\'onoma de M\'exico, Apdo. Postal 70-542, 04510 M\'exico, Distrito Federal, M\'exico}
\address{$^{(2)}$ Instituto de Ciencias F\'{\i}sicas, Universidad
  Nacional Aut\'onoma de M\'exico, Apdo. Postal 48-3, 
Cuernavaca, Morelos 62251, M\'exico}

\eads{\mailto{$^{(1)}$octavio.desantos@gmail.com}}

\begin{abstract}
Using the f-deformed oscillator formalism, we introduce two types of squeezed coherent states for a Morse potential system (Morse-like squeezed coherent states) through the following definitions: i) as approximate eigenstates of a linear combination of f-deformed ladder operators and ii) as deformed photon-subtracted coherent states. For the states thus constructed we analyze their statistical properties, their uncertainty relations, and their temporal evolution on phase space.  
\end{abstract}

\maketitle

\section{Introduction}

In recent times, the concept of coherent states has received a great deal of interest in many fields of physics, ranging from atomic and molecular physics to quantum optics. The coherent states of the quantum harmonic oscillator were first proposed by Schr\"odinger \cite{mandel}, in 1926, as states of minimum uncertainty product, being thus the quantum states which most closely resemble classical ones from the dynamical viewpoint; i.e., they remain well localized around their corresponding classical trajectory and fulfill the minimum uncertainty product for the position and momentum canonical variables, when the two standard variances are equal. After Schr\"odinger's proposal, almost four decades had to elapse before the notion of coherent states was widely recognized to be an important and convenient tool for the description of the quantum properties of the electromagnetic field, which was done by Glauber \cite{glauber}, in 1963, who first coined the term {\it coherent states.} More specifically, Glauber constructed what are now called {\it field coherent states} as eigenstates of the annihilation operator of the harmonic oscillator so as to study the correlation and coherence properties of the electromagnetic field, a theme of great relevance in quantum optics \cite{mandel,yamamoto}.  

Because of their increasing theoretical and experimental applicability in different aspects of modern physics \cite{klauderbook}, numerous generalizations of coherent states for systems other than the harmonic oscillator have been undertaken either from the dynamical point of view, from symmetry considerations, or from proposals based on deformed or nonlinear algebras. For instance, in terms of their evolution, it is worth mentioning the generalized coherent states constructed by Nieto and Simmons \cite{nieto1,nieto2,nieto3} for a number of confining one-dimensional systems, such as the harmonic oscillator with centripetal barrier and  symmetric P\"oschl-Teller potentials, among others. They were introduced as those states that minimize a generalized uncertainty relation in order that such states would imitate the classical motion in a given potential. On the other hand, Gazeau and Klauder \cite{gazeau} proposed another class of generalized coherent states for systems having a discrete and/or continuous spectrum, defining them in such a way that they possess continuity of labeling, a resolution of unity, and temporal stability. On the basis of symmetry considerations, coherent states for different Lie groups have been introduced. For example, we can cite the coherent states of the SU(1,1) Lie group constructed by Barut and Girardelo \cite{barut} as eigenstates of the ladder operators in such a representation. A general algorithm for constructing coherent states based upon the underlying symmetry associated with the system of interest was given, independently, by Gilmore \cite{gilmore1} and Perelomov \cite{perelomov}. Gilmore and Perelomov's definition generalizes the concept of coherent state in terms of a displaced reference state.

A nonlinear algebraic theory that has attracted much attention in recent years concerns a special quantum deformation process on the harmonic oscillator algebra, namely, the so-called {\it f-deformed oscillator formalism} proposed by Man'ko and co-workers \cite{manko1,manko2}. Within this algebraic scheme it turns out to be straightforward to construct the corresponding {f-coherent states} ({\it nonlinear coherent states}) as eigenstates of a deformed annihilation operator, whenever the appropriate deformation function that fixes the nonlinearity of the system under study is known. These states display nonclassical features like squeezing and anti-bunching effects, and they have also found interesting applications such as the description of quantum dots \cite{bagheri} and quantization of the center-of-mass motion of a laser-driven trapped ion \cite{matos,manko3,kis}. \\

Another class of quantum states that have attracted much attention in recent years, which will be the central theme of the this work, are the so-called squeezed coherent states. Besides being states of minimum uncertainty product, they are characterized by the property that one of the variances of two observables, either position or momentum of a mechanical oscillator or two orthogonal quadratures for a radiation field, is smaller than the vacuum (or coherent-state) noise level allowed by Heisenberg's uncertainty principle, at the expense of enhanced noise in the other observable \cite{dstoler,yuen}. Reduced fluctuations below the standard quantum limit in a given observable have found a manifold of applications in noise-free processes of measurement including spectroscopic techniques \cite{polzik}, optical interferometry \cite{xiao}, and gravitational wave detection \cite{caves}, among others.

The squeezed coherent states of the harmonic oscillator, also called two-photon coherent states, result from applying a unitary operator to the coherent state $|\alpha \rangle$ \cite{yuen}, i.e.,
\begin{equation}
|\xi,\alpha \rangle \equiv \hat{S}(\xi)|\alpha \rangle, 
\label{eq:scsho}
\end{equation}
where $\hat{S}(\xi)$, known as the {\it squeeze operator}, is defined as: 
\begin{equation}
\hat{S}(z) = \exp \left [\frac{1}{2}(\xi^{\ast}\hat{a}^{2}-\xi \hat{a}^{\dagger 2})\right ],
\end{equation}
with $ \xi=re^{i\theta}$. Indeed, the squeeze operator acting upon the coherent state reduces the dispersion of any one of the $\hat{x}$ and $\hat{p}$ conjugate variables of an electromagnetic field mode (or of its mechanical analog, a harmonic oscillator) in such a way that
\begin{eqnarray}
\langle \xi,\alpha | (\Delta \hat{x})^{2} |\xi,\alpha \rangle & = &  e^{-2r} \langle \alpha | (\Delta \hat{x})^{2} |\alpha \rangle, \\
 \langle \xi,\alpha | (\Delta \hat{p})^{2} |\xi,\alpha \rangle & = &  e^{2r} \langle \alpha | (\Delta \hat{p})^{2} |\alpha \rangle,
\end{eqnarray}
with $r$ being the {\it squeeze parameter}, while the uncertainty product takes on its smallest value
\begin{equation}
\langle \xi,\alpha | (\Delta \hat{x})^{2} |\xi,\alpha \rangle \langle \xi,\alpha | (\Delta \hat{p})^{2} |\xi,\alpha \rangle = \langle \alpha | (\Delta \hat{x})^{2} |\alpha \rangle \langle \alpha | (\Delta \hat{p})^{2} |\alpha \rangle = \frac{1}{4}.
\end{equation}

There exists a more general way of defining minimum-uncertainty squeezed states of any pair of physical quantities, and that is to build them as solutions of the eigenvalue equation 
\begin{equation}
(\hat{X}_{1}+i\lambda \hat{X}_{2}) |\psi \rangle = z |\psi \rangle,
\label{eq:eigenvalue}
\end{equation}
where $\hat{X}_{1}$ and $\hat{X}_{2}$ are noncommuting Hermitian operators in a given algebra, $\lambda$ is a generalized squeeze parameter, and $z$ a complex number. It is known that the $|\psi \rangle$ states satisfying the above equation, also known as intelligent states \cite{jackiw}, reduces the corresponding uncertainty relation to the equality
\begin{equation}
\langle (\Delta \hat{X}_{1})^{2} \rangle \langle (\Delta \hat{X}_{2})^{2} \rangle = \frac{1}{4} |\langle [\hat{X}_{1},\hat{X}_{2}] \rangle|^{2}.
\label{eq:unrel}
\end{equation}
It can be found in Ref. \cite{puri} that the squeezed coherent states defined by (\ref{eq:scsho}) indeed satisfy   eigenvalue equation (\ref{eq:eigenvalue}), on the condition that the operators $\hat{X}_{1}$ and $\hat{X}_{2}$ constitute hermitian generators of the harmonic oscillator algebra. Thus, by suitably redefining $z$ in terms of the parameters $\lambda$ and $\xi$ (see \cite{puri} for more details), such states may be viewed as eigenstates of a linear combination of the ladder operators $\hat{a}$ and $\hat{a}^{\dagger}$, i.e.,
\begin{equation}
\left [(1+\lambda)\hat{a}+(1-\lambda)\hat{a}^{\dagger} \right] |\psi \rangle = \sqrt{2} z|\psi \rangle,
\end{equation}
where the Hermitian operators $\hat{X}_{1}= \frac{1}{\sqrt{2}}(\hat{a}+\hat{a}^{\dagger})$ and $\hat{X}_{2}=\frac{1}{\sqrt{2}i}(\hat{a}-\hat{a}^{\dagger})$ are inserted into (\ref{eq:eigenvalue}). Further, it was shown in Ref. \cite{nieto4} that knowing the underlying dynamical algebra of the system under study one can extend the idea of employing linear combinations of ladder operators to cope with the problem of constructing squeezed states for general systems. Generalizations of squeezed coherent states for different algebras \cite{aragone,puriagarwal,fu,gerry1} and different potentials \cite{gerry2,hillery,bergou,kinani1,kinani2} have already been explored along this line. For instance, intelligent states associated with the SU(1,1) dynamical group and some of their realizations are reviewed in Ref. \cite{alkader}, where the corresponding eigenvalue problem to be solved is established by using the raising and lowering generators $\hat{K}_{\pm}$ for the said representation, namely,
\begin{equation}
\left [(1+\lambda)\hat{K}_{-}+(1-\lambda)\hat{K}_{+} \right] |\psi \rangle = 2z|\psi \rangle.
\label{eq:eigenvalue2}
\end{equation}
The analytic solution of the recurrence relation corresponding to this eigenvalue equation is expressed in terms of the Pollaczek polynomials. \\

Generalized minimum-uncertainty squeezed states related to nonlinear algebras, as well as some   considerations about how to perform their experimental realization as quantum states of the center-of-mass motion of a laser-driven trapped ion, have been introduced in \cite{shchukin}. More specifically, two eigenvalue problems for two general noncommuting operators, which are
\begin{eqnarray}
\left[ (1+\lambda)g(\hat{n})\hat{a}+(1-\lambda)\hat{a}^{\dagger}g(\hat{n}) \right] |\psi \rangle  =  z|\psi \rangle, \label{eq:gen1}\\  
\left [(1+\lambda)f(\hat{a})+(1-\lambda)f(\hat{a}^{\dagger}) \right] |\psi \rangle  =  z |\psi \rangle, \label{eq:gen2}
\end{eqnarray}
where both $g$ and $f$ are arbitrary functions, were solved by formulating its corresponding differential equation within the Fock-Bargmann representation. The first type of generalization (Eq. (\ref{eq:gen1})) yields what is called nonlinear squeezed states in the context of the quadratures of a deformed algebra, whereas the second one (Eq. (\ref{eq:gen2})) is a nonlinear generalization of the quadrature squeezed states.\\

A very recent and successful attempt to construct squeezed coherent states of a quantum system with a finite discrete spectrum modeled by the Morse potential was undertaken by Angelova \etal \cite{angelova}. To be more precise, by considering two different types of ladder operators, the authors were able to construct two different types of squeezed coherent states called oscillator-like and energy-like. These states turned out to be almost eigenstates of a linear combination of the ladder operators for the said potential, in the sense that they are solutions of the approximate eigenvalue equation:

\begin{equation}
(\hat{A}^{-}+\gamma \hat{A}^{+}) |\psi \rangle \approx z |\psi \rangle,
\end{equation}
with $\gamma$ being a squeezing parameter. Here, the ladder operators $\hat{A}^{-}$ and $\hat{A}^{+}$ satisfy either a su(2) or a su(1,1) algebra, each case leading to one group or the other of such states. \\

As far as the Morse oscillator is concerned, it is well-known that such a system has demonstrated to be a particularly useful anharmonic potential for the description of a little more realistic systems that deviate from the ideal harmonic oscillator case under certain conditions. In molecular physics, for instance, this model is widely used to represent the vibrational motion of a diatomic molecule approximated by a Morse potential. In nonlinear optics, on the other hand, it has been found that the discrete spectrum of certain nonlinear models such as the optical Kerr Hamiltonian (whose energy spectrum is unequally spaced) can be described as a Morse-like oscillator; in this sense, it is worth mentioning an interesting physical application introduced in Ref. \cite{chumakov} in which a Morse-type Hamiltonian with a finite spectrum is intended for modeling the properties of what is called a finite Kerr medium. Thus, because of its versatility, it is seen that the Morse-like oscillator can be used as a suitable algebraic model in different contexts. \\

Motivated by the aforementioned studies, throughout this work we shall consider two types of squeezed coherent states associated to the Morse potential (Morse-like squeezed coherent states) on the basis of the f-deformed oscillator formalism developed by Man'ko and coworkers \cite{manko1,manko2}. Within said formulation we choose a number-depending function $f(\hat{n})$ such that the energy spectrum of our deformed Hamiltonian model yields an energy spectrum similar to that of a Morse potential. Firstly, we adopt the ladder-operator technique outlined above for the construction of the first class of squeezed states as eigenstates of a linear combination of lowering and raising f-deformed operators $\hat{A}=\hat{a}f(\hat{n})$, $\hat{A}^{\dagger}=f(\hat{n})\hat{a}^{\dagger}$ written in terms of the usual harmonic oscillator creation and annihilation operators $\hat{a}$, $\hat{a}^{\dagger}$. Secondly, we proceed to introduce the second class of squeezed states in a not-so-conventional fashion, and that is to build them by iterative application of a f-deformed annihilation operator upon what is called a deformed displacement operator coherent state (the latter was previously constructed in Ref. \cite{osantos1} by application of a f-deformed displacement operator upon the ground state of the system). As such, these states are reminiscent of the so-called photon-depleted coherent states, for which a more detailed description will be given below. \\

This paper is organized as follows. In section 2 the Hamiltonian model that describes a Morse-like f-oscillator, the corresponding f-deformed ladder operators, and the two types of squeezed coherent states cited above are introduced. We devote section 3 to a wide discussion of the numerical results focused on the statistical properties, the phase space behavior, and the uncertainty relations of such states; in all calculations we constrain our approach to the low-lying region of the discrete energy spectrum. Finally, in section 4 some conclusions are given.

\section{The Morse-like f-deformed oscillator and its squeezed coherent states}

According to Man'ko \etal  \cite{manko1,manko2}, an f-oscillator is a non-harmonic quantum system  characterized by a Hamiltonian of the harmonic oscillator form

\begin{equation}
\hat{H}_{D} = \frac{\hbar \Omega}{2}(\hat{A}^{\dagger}\hat{A}+\hat{A}\hat{A}^{\dagger}),
\label{eq:defham1}
\end{equation}
where the deformed boson creation and annihilation operators $\hat{A}^{\dagger}$ and $\hat{A}$, respectively, are defined by deforming the standard harmonic operators $\hat{a}$ and $\hat{a}^{\dagger}$ via the non-canonical transformation
 \begin{equation}
\hat{A}=\hat{a} f(\hat{n}) = f(\hat{n}+1)\hat{a}, \qquad \hat{A}^{\dagger} = f(\hat{n}) \hat{a}^{\dagger} = \hat{a}^{\dagger} f(\hat{n}+1),
\label{eq:defops}
\end{equation}
where, in turn, $\hat{n}=\hat{a}^{\dagger} \hat{a}$ is the usual number operator, and the operator function $f(\hat{n})$, which is assumed to be real, is a deformation function depending on the level of excitation. 

From the deformed Hamiltonian (\ref{eq:defham1}), together with the canonical commutator $[\hat{a},\hat{a}^{\dagger}]=1$ and definitions (\ref{eq:defops}), an equivalent expression of it can be recast in terms of the number operator as follows

\begin{equation}
\hat{H}_{D}=\frac{\hbar \Omega}{2}((\hat{n}+1)f^{2}(\hat{n}+1)+\hat{n}f^{2}(\hat{n})).
\label{eq:defham2}
\end{equation}
Additionally, it is important to point out that the set of operators $\{\hat{A}, \hat{A}^{\dagger}, \hat{n} \}$ obeys the commutation relations
\begin{equation}
[\hat{A},\hat{n}] = \hat{A}, \qquad [\hat{A}^{\dagger}, \hat{n}] = -\hat{A}^{\dagger},
\label{eq:commutator1}
\end{equation}
and
\begin{equation}
[\hat{A},\hat{A}^{\dagger}] = (\hat{n}+1)f^{2}(\hat{n}+1)-\hat{n}f^{2}(\hat{n}).
\label{eq:commutator2}
\end{equation}

Unlike the case of the harmonic oscillator, it is clear that the commutator between the deformed operators $\hat{A}$ and $\hat{A}^{\dagger}$ is no longer a c-number, but it may become a rather complicated function of the number operator, depending on the particular choice of the deformation $f(\hat{n})$. Likewise, Hamiltonian (\ref{eq:defham2}) can no longer be, in general, a linear function of the number operator,  since additional powers of $\hat{n}$ may take place; needless to say, that explains why such systems are usually referred to as nonlinear. Finally, note that in the limit $f(\hat{n}) \to 1$ the harmonic oscillator algebra is completely recovered.\\

As we will see shortly, one of the advantages of using the f-deformed oscillator formalism is that one is able to choose a deformation function $f$ in such a way that the energy spectrum of the deformed Hamiltonian (\ref{eq:defham2}) yields the energy spectrum of the physical system that it seeks to describe. It is worth commenting that this framework has allowed us to describe the nonlinear nature of the modified and the trigonometric P\"oschl-Teller potentials, and construct their associated coherent states as well \cite{osantos2}. Thus, in the following we shall employ Hamiltonian (\ref{eq:defham2}) as an algebraic model so as to describe the Morse potential as a f-deformed quantum oscillator. \\

To illustrate the usefulness of the f-deformed algebra, let us choose the deformation function \cite{recamier1}
\begin{equation}
f^{2}(\hat{n}) = 1-\chi_{a} \hat{n},
\end{equation}
where $\chi_{a}$ is an anharmonicity parameter. In doing so, the deformed Hamiltonian becomes
\begin{equation}
\hat{H}_{D} = \hbar \Omega \left[\hat{n}+\frac{1}{2}-\chi_{a} \left(\hat{n}+\frac{1}{2} \right)^{2} -\frac{\chi_{a}}{4} \right].
\label{eq:spec1}
\end{equation}
As we can see, the spectrum of this Hamiltonian is in essence the same as that of the Morse potential \cite{landau}
\begin{equation}
E_{n} = \hbar \omega_{e} \left(n+\frac{1}{2} \right)-\frac{\hbar \omega_{e}}{2N+1} \left (n+\frac{1}{2}\right)^{2},
\label{eq:spec2}
\end{equation}
provided that $\omega_{e}=\Omega$ and $\chi_{a}=1/(2N+1),$ with $N$ being the number of bound states corresponding to the integers $0\le n \le N-1$. 

In addition, for this particular choice of deformation, it follows from Eq. (\ref{eq:commutator2}) that the commutator between $\hat{A}$ and $\hat{A}^{\dagger}$ is explicitly given by
\begin{equation}
[\hat{A}, \hat{A}^{\dagger}] = 1-\chi_{a}(2\hat{n}+1) = \frac{2N-2\hat{n}}{2N+1},
\label{eq:crdef}
\end{equation}
and the action of these operators has the following effect on the number operator sates $|n\rangle$:
\begin{eqnarray} 
\hat{A} |n\rangle & = &  \sqrt{n(1-\chi_{a}n)} |n-1\rangle, \nonumber \\
\hat{A}^{\dagger} |n\rangle & = &  \sqrt{(n+1)(1-\chi_{a}(n+1))} |n+1\rangle. \label{eq:acdef}
\end{eqnarray}
Note that these operators change the number of quanta in $\pm 1$ and their corresponding matrix elements are modified via the deformation function $f$. In other words, they create and annihilate a single anharmonic quantum within the context of a Morse-like potential system. Furthermore, from (\ref{eq:commutator1}) and (\ref{eq:crdef}), it is clear that this pair of deformed operators, together with the number operator, is closed under the operation of commutation, and, as such, they may be considered as the elements of the underlying dynamical algebra for the system being described. \\

Therefore, by virtue of the equivalence between (\ref{eq:spec1}) and (\ref{eq:spec2}), we shall employ the former as a suitable algebraic Hamiltonian in order to describe the discrete part of the  spectrum of a Morse system and evaluate the temporal evolution of its associated nonlinear  coherent states, provided we confine ourselves to the first $N$ bound states $\{ |0\rangle, |1\rangle, \ldots ,|N-1\rangle \}$ of such a system. Likewise, the nonlinear operators $\hat{A}$ and $\hat{A}^{\dagger}$ are considered to represent the deformed ladder operators that will enable us to define our first group of squeezed coherent states as usual, that is, as eigenstates of a linear combination of the dynamical variables associated with the system under study.

\subsection{Squeezed states defined as eigenstates of deformed ladder operators}

Having established the ladder operators associated with the Morse-like system, we set out to determine our first group of squeezed coherent states by the standard method. That is, let us define them as eigenstates of a linear combination of the deformed annihilation and creation operators as follows
\begin{equation}
(\hat{A}+\gamma \hat{A}^{\dagger}) |\alpha,\gamma \rangle = \alpha |\alpha, \gamma \rangle.
\label{eq:definition1}
\end{equation}
Here, the parameter $\gamma$ is a non-negative real number that takes on the role of a squeezing parameter, while $\alpha$ is called the {\it size} of the coherent state (it suffices for our purposes to assume $\alpha$ to be also real).

In order to solve Eq. (\ref{eq:definition1}) let the state $|\alpha,\gamma \rangle$ be expressed as a weighted superposition of number eigenstates
\begin{equation}
|\alpha, \gamma \rangle = \sum_{m=0} C_{m} |m\rangle,
\label{eq:sup1}
\end{equation}
where the summation must be truncated at $m=N-1$ for a Morse potential possessing $N$ bound states. Thereby, substitution of (\ref{eq:sup1}) in (\ref{eq:definition1}) gives the following approximate recursion relation for $C_{m}$
\begin{equation} \fl
C_{m+1} \sqrt{(m+1)(2N-m)}+\gamma C_{m-1} \sqrt{m(2N+1-m)} \approx \sqrt{2N+1} \alpha C_{m}.
\label{eq:recrel1}
\end{equation}
Now, we found it convenient to define
\begin{equation*}
\mathcal{C}_{m} = \frac{C_{m}}{\sqrt{m!(2N-m)!}}.
\end{equation*}
On inserting this in (\ref{eq:recrel1}) we get 
\begin{equation}
\mathcal{C}_{m+1}(m+1) +\gamma \mathcal{C}_{m-1} (2N-m+1) \approx \sqrt{2N+1} \alpha \mathcal{C}_{m}.
\label{eq:recrel2}
\end{equation}
This equation turns out to be an special case of an exactly solvable three-term recursion relation of the type (see Eq. (10.66) of Ref. \cite{puri})
\begin{equation}
m \alpha_{1}\mathcal{C}_{m}+(m+1)\gamma_{1}\mathcal{C}_{m+1}-\beta_{1} (M-m+1) \mathcal{C}_{m-1} = \lambda \mathcal{C}_{m},
\label{eq:recrel3}
\end{equation}
where $\alpha_{1}$, $\beta_{1}$ and $\gamma_{1}$ are fixed constants, $\lambda$ is an eigenvalue, $m=0,1, \ldots ,M$ ($M$ being a finite number describing the upper limit on the allowed values of $m$, i.e. $\mathcal{C}_{M+j}=0$ for $j \ge 1$), and $\mathcal{C}_{-1}=0$. Then, in accordance with Ref. \cite{puri} the solution of (\ref{eq:recrel3}) assumes the general form 
\begin{equation}
\mathcal{C}_{mn} = (-1)^{M+m} \sum_{k=0}^{n} {n \choose k} {M-n \choose m-k} x_{+}^{n-k} x_{-}^{M-n-m+k},
\label{eq:gensol}
\end{equation}
where
\begin{eqnarray}
x_{\pm} & = & \frac{1}{2\beta_{1}} \left [-\alpha_{1}\pm \sqrt{\alpha_{1}^{2}-4\gamma_{1}\beta_{1}}\right], \\
n & = & \frac{\lambda+M\beta_{1}x_{+}}{\beta_{1}(x_{+}-x_{-})}.
\end{eqnarray}
So, taking $\alpha_{1} =0$, $\gamma_{1}=1$, $\beta_{1}=-\gamma$, $M=2N$, and $\lambda = \sqrt{2N+1} \alpha$ in Eq. (\ref{eq:recrel3}), it is effortless to obtain the corresponding analytical solution of (\ref{eq:recrel2}) when referring to Eq. (\ref{eq:gensol}), which, under the normalization condition $\langle \alpha, \gamma |\alpha, \gamma \rangle = 1$, takes the form
\begin{eqnarray}
C_{mn} & = & N_{n} \sqrt{m! (2N-m)!} \mathcal{C}_{nm} \nonumber \\ 
& = &  N_{n} (-1)^{m} \gamma^{m/2} \sqrt{m! (2N-m)!} \sum_{k=0}^{n} {n \choose k} {2N-n \choose m-k}(-1)^{k},
\label{eq:solution}
\end{eqnarray}
with $N_{n}$ being a normalization constant given by
\begin{equation}
N_{n} =\left  \{ \sum_{m=0}^{N-1} \gamma^{m} m! (2N-m)! \left( \sum_{k=0}^{n} {n \choose k} {2N-n \choose m-k}(-1)^{k} \right )^{2} \right \}^{-1/2},
\end{equation}
where, in turn, $n=N+\left[\frac{\alpha}{2\sqrt{\chi_{a}\gamma}} \right]$; here, the notation $[x]$ denotes the operation of taking the integer closest to $x$. Finally, on substituting (\ref{eq:solution}) in (\ref{eq:sup1}), our first group of Morse-like squeezed coherent states is explicitly written as 

\begin{equation}
\fl |\alpha, \gamma \rangle = N_{n} \sum_{m=0}^{N-1} (-1)^{m} \gamma^{m/2} \sqrt{m! (2N-m)!} \sum_{k=0}^{n} {n \choose k} {2N-n \choose m-k}(-1)^{k} |m \rangle.
\label{eq:edos}
\end{equation}

By comparing Eqs. (\ref{eq:recrel2}) and (\ref{eq:recrel3}) one can see that Eq. (\ref{eq:solution}) is in fact a solution of $2N+1$ allowed elements of $m$. Nevertheless, only the first $N$ elements will be the ones of physical interest to us, since we are dealing with a Morse potential containing such a number of bound states, as stated before. For this approximation to be valid, it is important to be careful in using suitable values for $\alpha$ and/or $\gamma$ so as to constrain the evolution of our coherent states to the low-lying region of the spectrum. The states thus obtained will be called deformed ladder-operator quasi-coherent states (LOQCSs).

\subsection{Deformed photon-subtracted coherent states}

First of all, let us start by recalling that within the context of quantum optics there exists an interesting family of nonclassical states emerging from the process of adding $m$ quanta (photons) to a given quantum state. The generic form of these states reads as
\begin{equation}
|\psi, m \rangle = N_{m} \hat{a}^{\dagger m}| \psi \rangle,
\label{eq:defpa}
\end{equation}
where $|\psi \rangle$ denotes an arbitrary initial state, $\hat{a}^{\dagger}$ is the standard creation operator, $m$ is taken to be an non-negative integer (the number of added quanta), and $N_{m}$ is a normalization constant. Of particular interest are those states for which the initial state $|\psi \rangle$ is chosen to be a harmonic oscillator coherent state; we refer to the {\it photon-added coherent states} introduced for the first time by Agarwal and Tara \cite{agarwaltara}. Much work has been devoted to the study of these states, both in their nonclassical properties such as squeezing and sub-Poissonian statistics (see \cite{dodonov} and the references therein contained) and in putting forward some prospects for their physical realization \cite{dakna,kalamidas,zavatta,sivakumar}. For instance, it was first proposed in Ref. \cite{agarwaltara} that the photon-added states may be produced in the interaction between a two-level atom and a single-mode cavity field. More precisely, the authors used as starting point a time independent Hamiltonian of the form
\begin{equation}
\hat{H} = \hbar \eta (\hat{\sigma}_{+} \hat{a}+\hat{\sigma}_{-}\hat{a}^{\dagger}),
\label{eq:hamint}
\end{equation}
where $\hat{a}$ and $\hat{a}^{\dagger}$ are the annihilation and creation operators for the field, respectively, while $\hat{\sigma}_{+}$ and $\hat{\sigma}_{-}$  are the standard atomic two-level transition operators acting upon the levels $|g\rangle$ (ground state) and $|e\rangle$ (excited state) of the atom in such a way that $|g \rangle \stackrel{\hat{\sigma}_{+}} {\longrightarrow} |e\rangle$ and $|e \rangle \stackrel{\hat{\sigma}_{-}} {\longrightarrow} |g\rangle$. So, if the initial state of the atom-field system is $|\alpha \rangle |e\rangle$, where $|\alpha \rangle$ is a coherent state of the field, then the state $|\psi \rangle$ of the composite system at time $t$ is given by
\begin{equation}
|\psi (t) \rangle = \exp (-i\hat{H}t/\hbar) |\alpha \rangle |e\rangle.
\end{equation}
In addition, if the coupling constant $\eta$ is small enough, and the interaction duration is sufficiently short so that $\eta t \ll 1$, it is allowed to rewrite the above equation as
\begin{equation}
|\psi (t) \rangle \approx |\alpha \rangle |e\rangle -\frac{iHt}{\hbar} |\alpha \rangle |e\rangle.
\label{eq:finalstate1}
\end{equation}
Finally, substitution of (\ref{eq:hamint}) in (\ref{eq:finalstate1}) gives 
\begin{equation}
|\psi (t) \rangle \approx |\alpha \rangle |e\rangle -i \eta \hat{a}^{\dagger} |\alpha \rangle |g\rangle.
\label{eq:finalstate2}
\end{equation}
This shows that it is, in principle, possible to produce what is referred to as a single-photon-added coherent state, $\hat{a}^{\dagger}|\alpha \rangle$, if the atom is detected to be in the ground state $|g\rangle$. The same approach holds for the case of multiphoton-excitation states proportional to $\hat{a}^{\dagger m}|\alpha \rangle$ in doing the replacement $\hat{a}(\hat{a}^{\dagger})$ by $\hat{a}^{m}(\hat{a}^{\dagger m})$ into Eq. (\ref{eq:hamint}).

It is worth pointing out that there has also been some interest focused on the study of possible generalizations of photon-added coherent states for systems other than the harmonic oscillator. As an example, let us mention the {\it deformed photon-added nonlinear coherent states} (DPANCSs) which are defined as \cite{safaeian}
\begin{equation}
|\alpha,f,m\rangle = N_{\alpha}^{m}\hat{A}^{\dagger m} |\alpha,f\rangle,
\end{equation}
where, as seen from Eq. (\ref{eq:defpa}), the standard creation operator is replaced by its deformed counterpart $\hat{A}^{\dagger}$, the initial state is regarded as an eigenstate of the deformed annihilation operator $\hat{A}$ (i.e., a nonlinear coherent state defined via $\hat{A}|\alpha,f\rangle = \alpha |\alpha,f\rangle$), and $N_{\alpha}^{m}$ is a normalization coefficient to be determined. Moreover, in order to produce this nonlinear version of photon-added states in similarity to the harmonic oscillator case, it was proposed by the authors of \cite{safaeian} an interaction Hamiltonian to be written in terms of the deformed operators, that is, a Hamiltonian of the type
\begin{equation}
\hat{H} = \hbar \eta (\hat{\sigma}_{+} \hat{A}+\hat{\sigma}_{-}\hat{A}^{\dagger}),
\label{eq:hamintdef}
\end{equation}
where, as we can see, its essential characteristic lies on an intensity-dependent coupling between the atom and the cavity field modeled by the deformation function. By proceeding in this way, and choosing $|\alpha,f\rangle |e\rangle$ as the initial state of the composite system, one arrives at
\begin{equation}
|\psi (t) \rangle \approx |\alpha,f \rangle |e\rangle -i \eta \hat{A}^{\dagger} |\alpha,f \rangle |g\rangle,
\end{equation}
which leads clearly to a deformed single-photon-added coherent state, $\hat{A}^{\dagger}|\alpha,f\rangle$, if the atom is measured to be in the ground state. The same reasoning applies to multiphoton processes via the interaction Hamiltonian $\hat{H} = \hbar \eta (\hat{\sigma}_{+} \hat{A}^{m}+\hat{\sigma}_{-}\hat{A}^{\dagger m})$ \cite{safaeian}. \\

In this work we are interested in analyzing the converse process of that outlined above. We will examine the properties of a class of states emerging from the process of {\it subtracting} $m$ quanta from a given state. So then, let us consider the case wherein the creation operator $\hat{a}^{\dagger}$ in definition (\ref{eq:defpa}) is replaced by the deformed annihilation operator $\hat{A}$ as follows 
\begin{equation}
|\psi, m \rangle = N_{m} \hat{A}^{m} | \psi \rangle.
\label{eq:defps1}
\end{equation}
Certainly, if one chooses the initial state to be the nonlinear coherent state $|\alpha,f\rangle$, then it can easily be inferred that the definition leaves the state $|\psi, m \rangle$ in the coherent state $|\alpha,f \rangle$ itself, which is not useful for our purposes. Instead, in order to circumvent this minor problem, we shall employ a different type of initial state. \\

For a Morse-like oscillator, it was found in Ref. \cite{osantos1} that after applying the corresponding f-deformed displacement operator upon the ground state of the system, the resulting deformed coherent states appear to be nearly well-localized on the quantum phase space from the viewpoint of their Wigner distribution function. Such states are defined explicitly as
\begin{equation}
|\zeta \rangle \equiv \hat{D}(\zeta(\alpha)) |0\rangle \approx \frac{1}{(1+|\zeta|^{2})^{N}} \sum_{n=0}^{N-1} {2N \choose n}^{1/2} \zeta^{n} |n\rangle,
\label{eq:docs}
\end{equation}
where the deformed displacement operator $\hat{D}(\zeta(\alpha))=\exp(\alpha \hat{A}^{\dagger}-\alpha^{\ast} \hat{A}) $ is constructed as a generalization of the usual displacement operator $\hat{D}(\alpha) = \exp(\alpha \hat{a}^{\dagger}-\alpha^{\ast} \hat{a})$; the problem of disentangling the exponential was overcome by making use of Lie algebraic methods \cite{puri}. Here, given that we are dealing with a system having a finite number $N$ of bound states, the summation in (\ref{eq:docs}) must end at $n=N-1$, reason why these states are only approximate. For a given value of $\alpha=|\alpha|e^{i\phi}$, it has also been introduced the new complex parameter $\zeta = e^{i \phi}\tan(|\alpha|\chi_{a})$. Unlike the harmonic oscillator case, note that this type of nonlinear coherent states are not eigenstates of the deformed annihilation operator. They are referred to as {\it deformed displacement operator coherent states} (DOCS) \cite{osantos1}. \\

By using the DOCS (\ref{eq:docs}) as the initial state $|\psi \rangle$ into the definition (\ref{eq:defps1}), one gets the desired result: 
\begin{equation}
\fl  |\zeta, m\rangle = \hat{A}^{m} \hat{D}(\zeta) |0\rangle = \frac{C_{\zeta,m}}{(1+|\zeta|^{2})^{N}} \sum_{n=0}^{N-1-m} {2N \choose n+m}{2N \choose n}^{-1/2} \frac{(n+m)!}{n!} \zeta^{n+m} |n\rangle,
\label{eq:defps2}
\end{equation}
where
\begin{equation}
\fl C_{\zeta,m} = (1+|\zeta|^{2})^{N} \left (\sum_{n=0}^{N-1-m} {2N \choose n+m}^{2}{2N \choose n}^{-1} \left (\frac{(n+m)!}{n!} \right)^{2} |\zeta|^{2(n+m)} \right )^{-1/2}
\end{equation}
is a normalization coefficient. We refer to these states as the {\it deformed photon-subtracted coherent states} (DPSCSs). 

We have thus obtained a class of quantum states that evokes a type of photon-depleted coherent states introduced in the literature. Those multiphoton-annihilation states have been constructed by making use of inverse-ladder-operator techniques within the context of both the q-deformed and generalized hypergeometric coherent states; for more details see Refs. \cite{naderi,houn}. However, because of their definition, we consider that our photon-subtracted states are more likely to be realized in multiphoton processes produced in nonlinear media. Just for justifying this perception, let us return to the atom-field nonlinear interaction governed by Hamiltonian (\ref{eq:hamintdef}). Let $|\zeta \rangle |g\rangle$ be the initial state of the atom-field system; i.e., at the beginning of the interaction, it is assumed the photon-number probability density of the field to be the one of the DOCS defined in (\ref{eq:docs}), with the atom being in its ground state. Then, for short interaction times, the atom-field state becomes 
\begin{equation}
|\psi (t)\rangle \approx |\zeta \rangle |g\rangle -i\eta t \hat{A} |\zeta \rangle |e \rangle.
\end{equation}
It is clear that we will find the state of the field in the deformed single-photon-subtracted coherent state $\hat{A}|\zeta \rangle$ whether the atom is detected to be in the excited state $|e\rangle$ (a higher-order process would be required to deplete a greater number $m$ of quanta from the initial coherent state). Thus, it makes sense to consider the definition (\ref{eq:defps1}), together with the displacement operator states (\ref{eq:docs}), as an alternative way of constructing photon-depleted states. \\

We close this section with the following remarks. As already mentioned, the particular scheme for the generation of photon-depleted-like states outlined above is based on a quantum process whereby the initial state of the system is considered to be a deformed state of the radiation field in terms of the photon number statistics. The use of non-conventional coherent states for the description of the statistical properties of the field should not be surprising anymore. There have been a variety of proposals intended for modeling nonclassical light or non-ideal lasers by means of deformed photon states under physically realistic conditions in which deviations from the Poissonian distribution may take place. In this regard, we can quote the application of q-deformed coherent states to the characterization of the photon statistics of non-conventional light exhibiting either sub-Poissonian or super-Poissonian statistics \cite{katriel}. It is also important to mention the so-called binomial coherent states introduced by Stoler \etal \cite{stoler} for which coefficients of their photon-counting probability distribution is defined to be binomial in a finite-dimensional space. 

On the other hand, from a mechanical point of view, it has been shown in Refs. \cite{wallentowitz,matos2} that the vibrational coupling of a trapped atom by appropriate laser excitations allows us to study phenomena we are familiar with from quantum optics, such as Kerr-like effects, parametric interactions, and several types of nonlinear wave mixings. It turns out that, in certain instances, effective Hamiltonians of type (\ref{eq:hamintdef}) may be realized within the context of the quantized motion of a single atom, where the trap potential would imitate a cavity used in quantum optics (see, for example, Eq. (1) of Ref. \cite{matos2}). So, in light of these studies, we consider that it is, in principle, possible to generate photon-depleted-type coherent states which can be viewed in terms of annihilations of motional quanta. Along this sort of line, a class of photon-added coherent states in the case of a time-dependent frequency of the oscillator was analyzed in Ref. \cite{dodonov}. According to the authors, such a case can be thought of as a time dependence of the frequency of the electromagnetic field in a cavity (which may be ascribable to the motion of the cavity walls) or a vibrational frequency of an ion inside an electromagnetic trap. Given this, the concept of photon-subtracted-type (or photon-added-type) states makes sense in both a mechanical and optical context.

\section{Numerical results}

We are now in a position to proceed to the calculation of some physical quantities of interest related to the Morse-like squeezed coherent states obtained above. To begin with, we examine their statistical properties via the normalized variance of the number operator. Next, we analyze their temporal behavior on phase space and dispersion relations. From now on, we shall consider a Morse-like oscillator possessing $N=10$ ($\chi_{a}\approx 0.05 $) bound states. 

\subsection{Statistical properties}

Deviations of the occupation number distribution from a Poissonian statistics (which, as known, is characteristic of the harmonic oscillator coherent states) can be evaluated by means of the normalized variance of the number operator \cite{schleich}
\begin{equation}
\frac{\langle (\Delta \hat{n})^{2}\rangle}{\langle \hat{n} \rangle} = \frac{\langle \hat{n}^{2} \rangle -\langle \hat{n} \rangle^{2}}{\langle \hat{n} \rangle},
\label{eq:norvar}
\end{equation}
where the average values are taken between either the deformed ladder-operator quasi-coherent states (LOQCS) or the deformed photon-subtracted coherent states (DPSCS). Values of this quantity such that $\langle (\Delta \hat{n})^{2}\rangle /\langle \hat{n} \rangle <1$ or $\langle (\Delta \hat{n})^{2}\rangle /\langle \hat{n} \rangle > 1$ correspond to sub-Poissonian or super-Poissonian statistics, respectively; the former is known to be a nonclassical feature \cite{davidovich}.  Poissonian distribution means that $\langle (\Delta \hat{n})^{2}\rangle /\langle \hat{n} \rangle = 1$. \\

\begin{figure}[t!]
\begin{center}
\includegraphics[width=12cm, height=4cm]{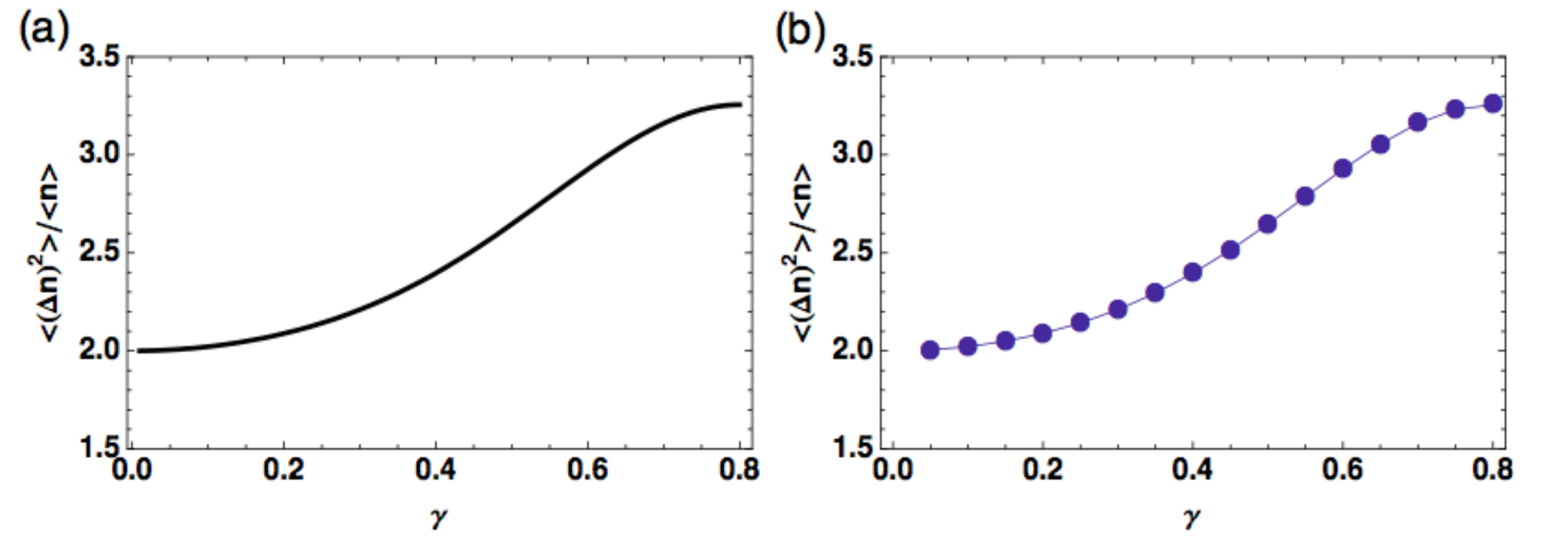} 
\caption{Plot of the normalized variance of the number operator, $\langle (\Delta \hat{n})^{2}\rangle /\langle \hat{n} \rangle =(\langle \hat{n}^{2} \rangle -\langle \hat{n} \rangle^{2})/\langle \hat{n} \rangle$, as a function of $\gamma$ for the Morse-like squeezed vacuum state $|\alpha=0,\gamma \rangle$ (frame (a)) compared with the numerical solution of Eq. (\ref{eq:recrel1}) (frame (b)).}
\label{fig:statistics1}
\end{center}
\end{figure}

\begin{figure}[t!]
\begin{center}
\includegraphics[width=12cm, height=8cm]{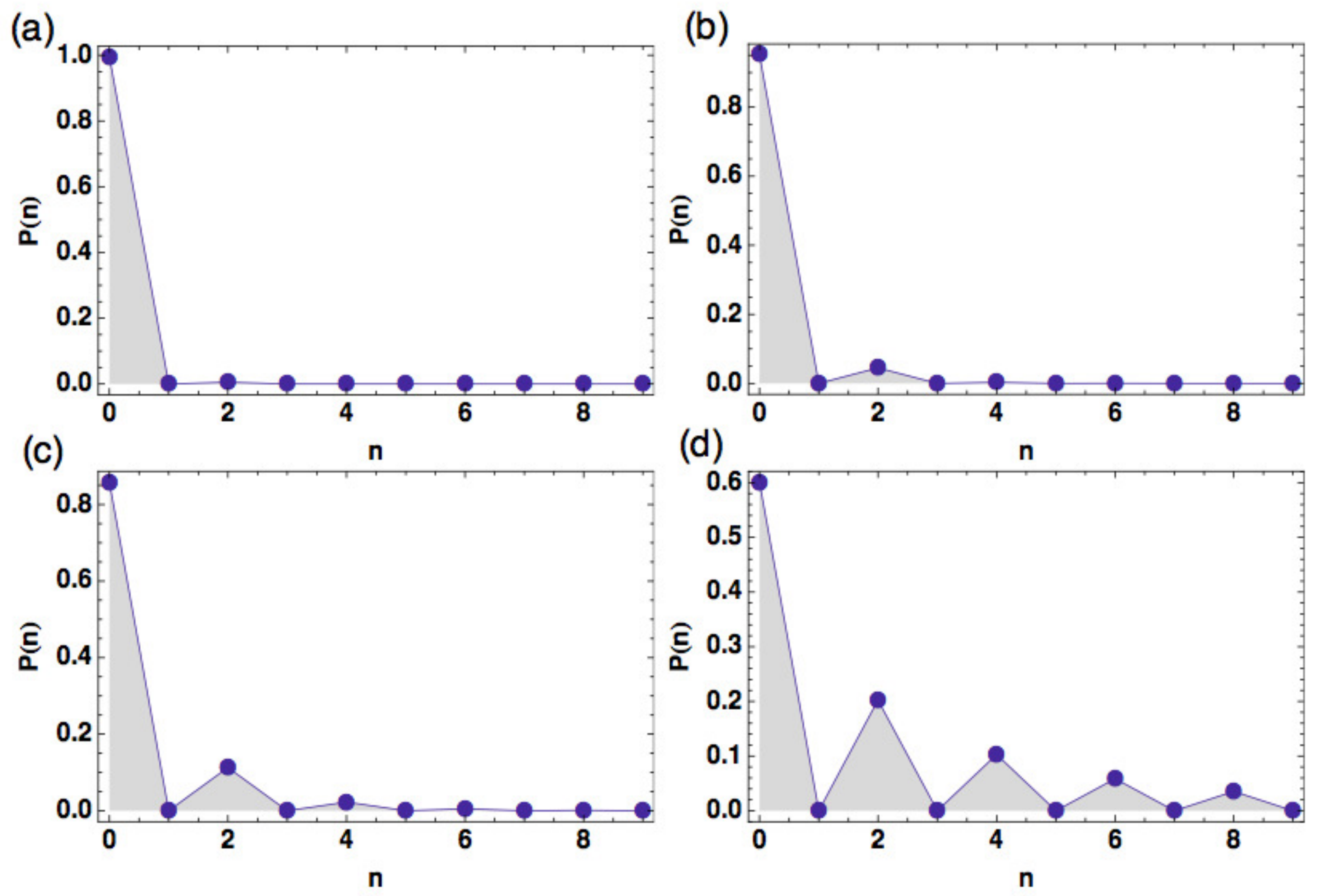} 
\caption{Plots of the number-state probability distribution for the Morse-like squeezed vacuum state $|\alpha=0,\gamma \rangle$. Frames (a) to (d) correspond to $\gamma= 0.1$, $0.3$, $0.5$ and $0.8$, respectively.}
\label{fig:statistics2}
\end{center}
\end{figure}

\begin{figure}[h!]
\begin{center}
\includegraphics[width=12cm, height=8cm]{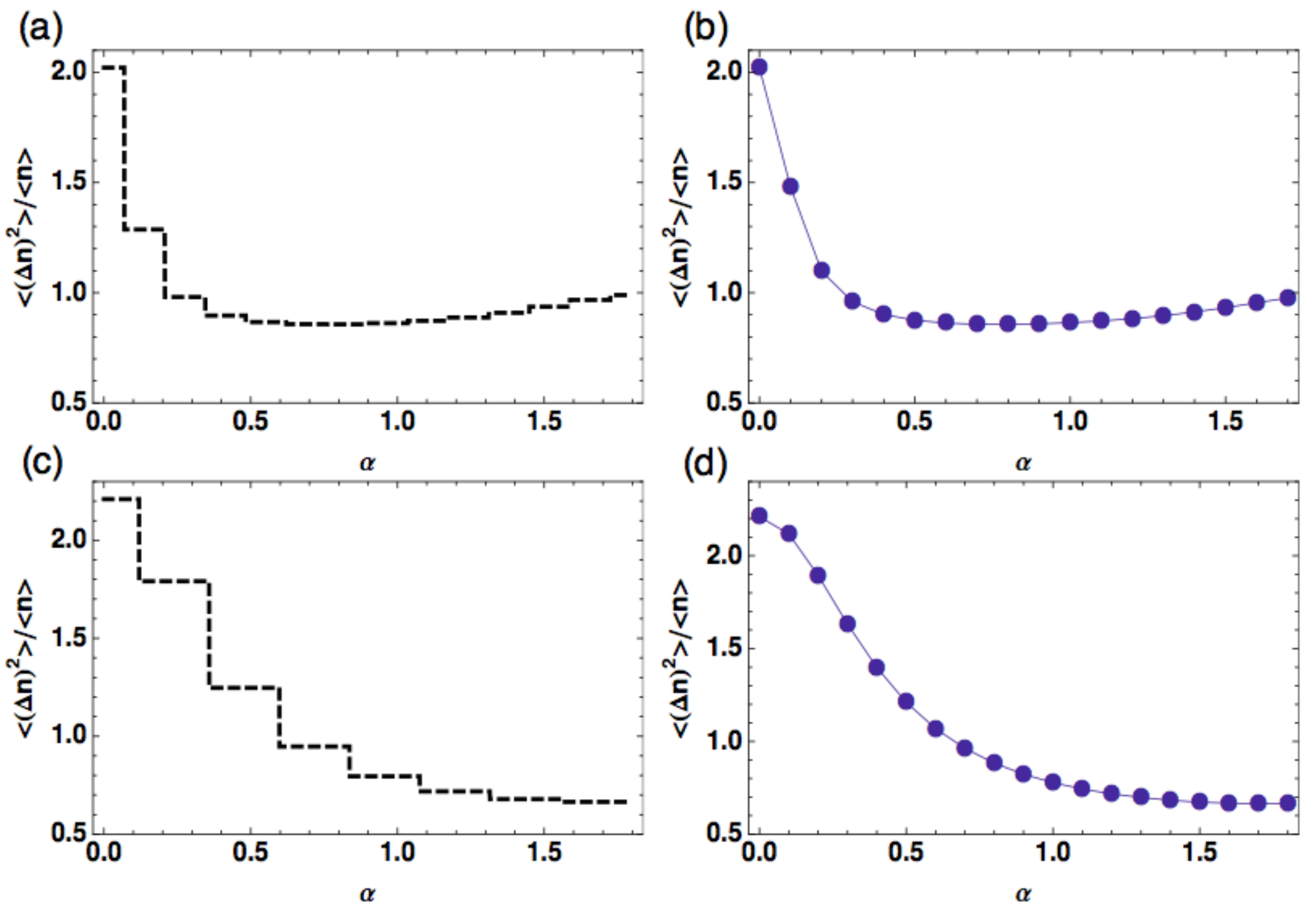} 
\caption{Plots of the normalized variance of the number operator, $\langle (\Delta \hat{n})^{2}\rangle /\langle \hat{n} \rangle =(\langle \hat{n}^{2} \rangle -\langle \hat{n} \rangle^{2})/\langle \hat{n} \rangle$, as a function of $\alpha$ for Morse-like squeezed states with $\gamma =0.1$ (frames (a), (b)) and $\gamma = 0.3$ (frames (c), (d)). Results appearing in frames (a) and (c) were obtained by using the analytical expression of $|\alpha,\gamma \rangle$ defined in (\ref{eq:edos}), while those results displayed in frames (b) and (d) come from solving Eq. (\ref{eq:recrel1}) numerically.}
\label{fig:statistics3}
\end{center}
\end{figure}

\begin{figure}[h!]
\begin{center}
\includegraphics[width=12cm, height=8cm]{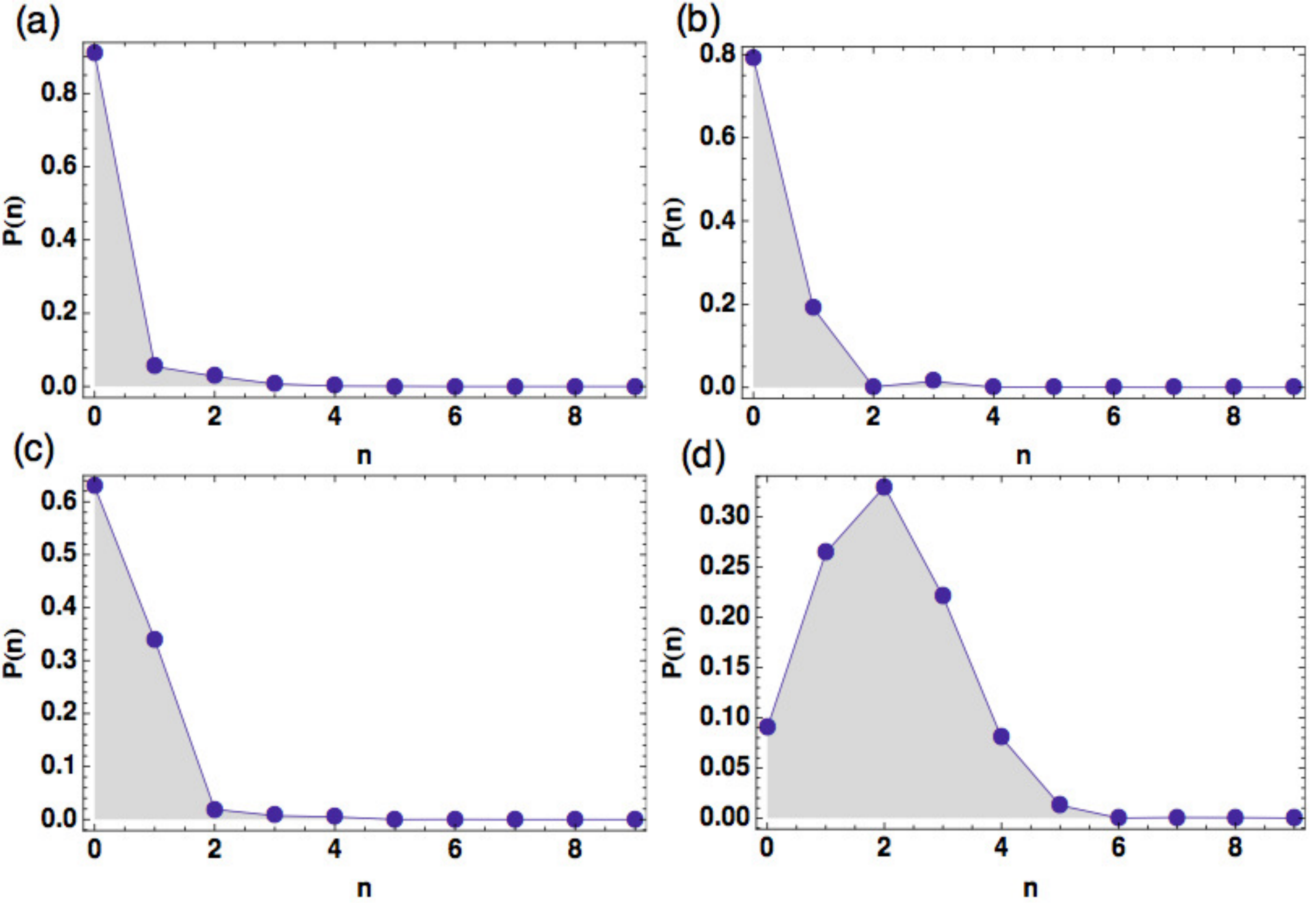} 
\caption{Plots of the number-state probability distribution for Morse-like squeezed coherent states with $\gamma=0.3$. Frames (a) to (d) correspond to $\alpha= 0.2$, $0.4$, $0.8$ and $1.6$, respectively.}
\label{fig:statistics4}
\end{center}
\end{figure}

Figure \ref{fig:statistics1} shows the normalized variance, Eq. (\ref{eq:norvar}), as a function of the squeezing parameter $\gamma$ for the Morse-like squeezed vacuum state which corresponds to the particular case of the LOQCS, defined by Eq. (\ref{eq:edos}), with $\alpha =0$. In order to confirm the feasibility of this analytic expression, we have added in Fig. \ref{fig:statistics1} (b) the result from the numerical solution of the recursion relation given by Eq. (\ref{eq:recrel1}). Both calculations lead to the same statistical behavior as expected. We can see that such a behavior is essentially super-Poissonian ($\langle (\Delta \hat{n})^{2}\rangle /\langle \hat{n} \rangle > 1$) within  the interval $0<\gamma \le 0.8$. Its ascending profile is closely related to spreading effects in the number-occupation probability distribution associated with such states. For example, let us consider the probability that the state $|\alpha=0,\gamma \rangle$ has the excitation number $n$, namely, $p_{n}= |\langle n|0,\gamma \rangle|^{2}$ evaluated by using (\ref{eq:edos}). In Fig. \ref{fig:statistics2} we have plotted $p_{n}$ as a function of $n$ for $\gamma=0.1$, $0.3$, $0.5$, and $0.8$. We see that as the squeeze parameter $\gamma$ increases, the occupation number distribution $p_{n}$ becomes wider, which is consistent with the fact that the dispersion in the number operator $\hat{n}$ is larger than its mean.  And, in conjunction with this tendency, $p_{n}$ reveals an oscillatory behavior, being zero for odd $n$. On the other hand, Fig. \ref{fig:statistics3} shows the more general case for which the normalized variance of the LOQSCs is plotted as a function of $\alpha$ for $\gamma =0.1$ and $0.3$. Clearly, in such a case the statistical behavior seems very different to the one displayed in Fig. \ref{fig:statistics1}, ranging from a super-Poissonian to sub-Poissonian conduct; the plots appearing in Fig. \ref{fig:statistics3}(a) and (c) owe their stepped shape to the particular way in which the LOQCSs are algebraically constructed (see Eq. (\ref{eq:edos})). Again, the corresponding numerical results calculated from Eq. (\ref{eq:recrel1}) (see Figs. \ref{fig:statistics3}(b) and (d)) yield a confirmation of the validity of Eq. (\ref{eq:edos}). In Fig. \ref{fig:statistics4}, we have also plotted $p_{n}$ for the coherent state $|\alpha,\gamma=0.3\rangle $ (Eq. (\ref{eq:edos})) as a function of $n$ for $\alpha =0.2$, $0.4$, $0.8$ and $1.6$. It is found that the occupation number distribution is less oscillatory for small values of $\alpha$, approximately for $0.2<\alpha<0.5$. Such a characteristic is almost vanished for $\alpha \ge 0.5$. Note that we have focussed our attention on those energy regions where the influence of the continuum can be discarded. \\

\begin{figure}[h!]
\begin{center}
\includegraphics[width=12cm, height=4cm]{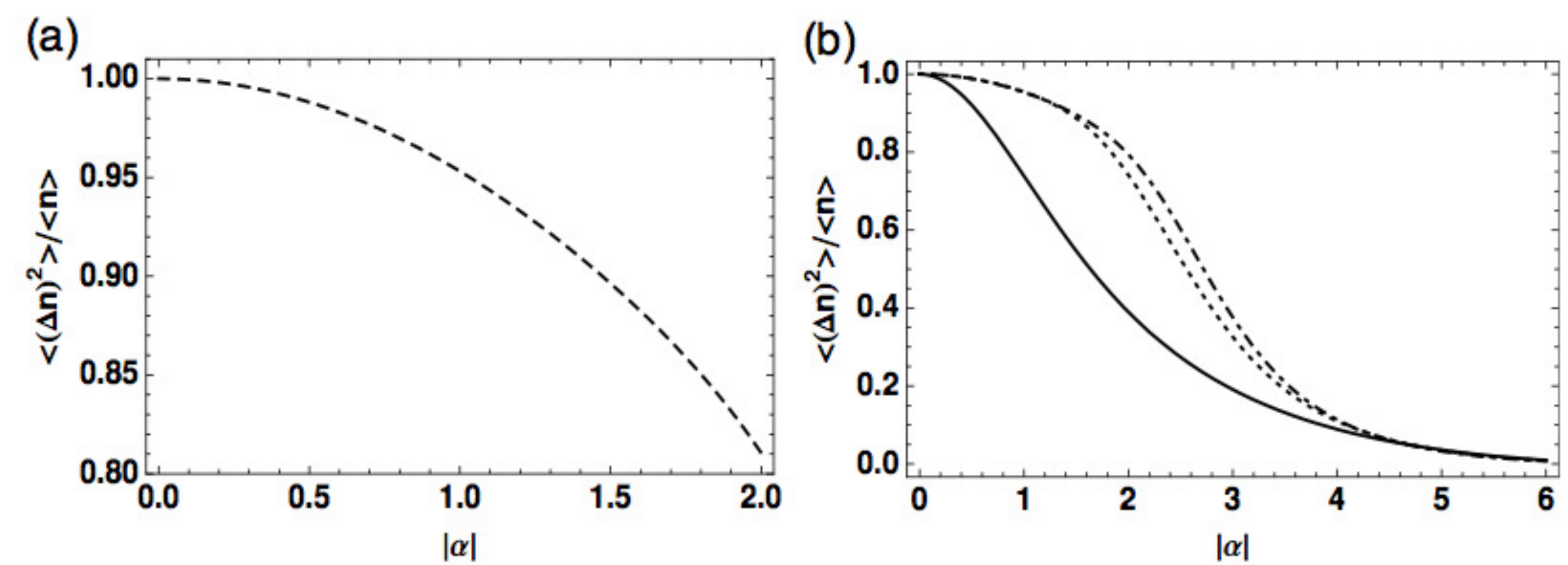} 
\caption{Plots of the normalized variance of the number operator, $\langle (\Delta \hat{n})^{2}\rangle /\langle \hat{n} \rangle =(\langle \hat{n}^{2} \rangle -\langle \hat{n} \rangle^{2})/\langle \hat{n} \rangle$, as a function of $\alpha$ for photon-subtracted coherent states $|\zeta,m \rangle$ with $m=0$ (dashed line), $m=2$ (dot-dashed line), $4$ (dotted line), and $8$ (continuous curve).}
\label{fig:statistics5}
\end{center}
\end{figure}

\begin{figure}[h!]
\begin{center}
\includegraphics[width=12cm, height=8cm]{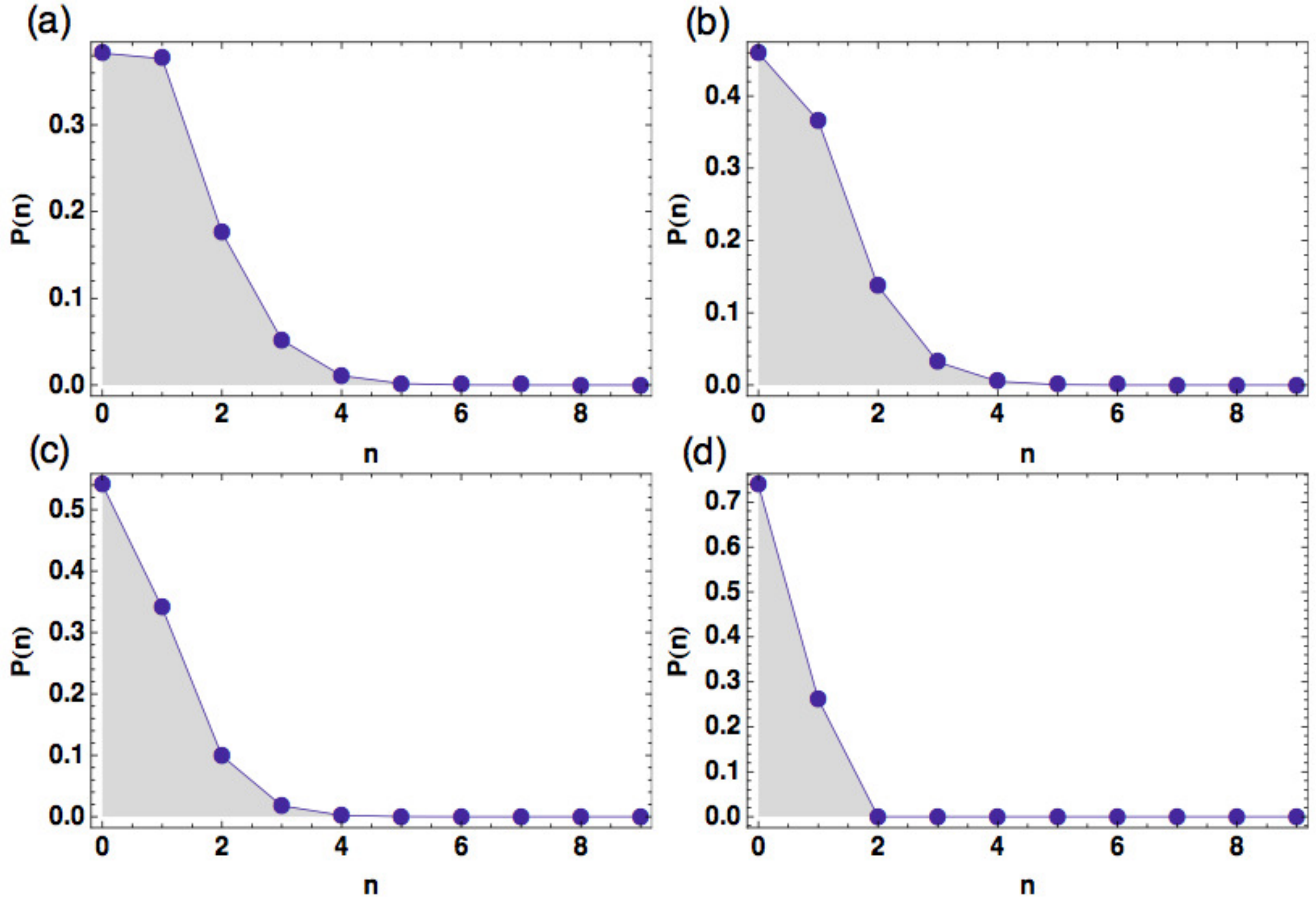} 
\caption{Plots of the number-state probability distribution for photon-subtracted coherent states with $\alpha=1$. Frames (a) to (d) correspond to $m= 0$, $2$, $4$ and $8$, respectively.}
\label{fig:statistics6}
\end{center}
\end{figure}

For the DPSCSs defined by Eq. (\ref{eq:defps2}), the normalized variance of the number operator turns out to be  a decreasing function of $\alpha$, as seen in Fig. \ref{fig:statistics5}. Frame \ref{fig:statistics5}(a) shows the particular case $m=0$ corresponding to the DOCS defined in (\ref{eq:docs}), where the $\alpha$ range shown is $0\le |\alpha| \le 2$. Although in the limit $|\alpha| \to 0$ the statistical behavior of the DPSCSs tends to be nearly Poissonian, in essence they exhibit a sub-Poissonian character as the value of $|\alpha|$ increases, which is better visualized in frame \ref{fig:statistics5}(b) where the plots were obtained from subtracting $m=2$, $4$, and $8$ quanta from the deformed displacement operator coherent state. It is found that the descending feature of these graphs becomes more significant insofar as the values of both $m$ and $|\alpha|$ increase. For the said values of $m$, we have shown in Fig. \ref{fig:statistics6} the corresponding occupation number probability distribution $p_{n}=|\langle n|\zeta(\alpha),m\rangle|^{2}$ for the DPSCSs with $\alpha =1$. In this picture we note that the successive depletion of quanta from the initial state appears to be a little more evident, particularly for $m\ge 4$, and herein lies the sub-Poissonian conduct that we have seen in Fig. \ref{fig:statistics5} for such states.

\subsection{Phase space and uncertainty relations}

In order to compute the temporal evolution on phase space of the coherent states so far obtained, as well as their dispersion and uncertainty relations, we will make use of a convenient algebraic representation for the position and momentum operators put forward by Carvajal \etal \cite{carvajal} in their work on the SU(2) model of vibrational excitations for Morse oscillators. To do this, we only need to take into consideration the following approximations:
 
First, if the number of bound states supported by the system is such that $2N \gg 1$, then the commutator between the deformed operators (\ref{eq:crdef}) may be approximated by
\begin{equation}
[\hat{A}, \hat{A}^{\dagger}]  \approx 1-\frac{\hat{n}}{N},
\end{equation}
and the commutation relationships (\ref{eq:commutator1}) turn out to coincide with those obtained in Ref. \cite{carvajal}, in which case a one-to-one correspondence between both formulations can be established. If so, the position and momentum variables may be given as a series expansion involving all powers of the deformed creation and annihilation operators; for more details see \cite{carvajal}. Next, since we are focusing our attention on the low-lying region of the spectrum, we shall employ an approximate version of such a representation only up to second order in $\hat{A}$ and $\hat{A}^{\dagger}$, that is,
{\setlength\arraycolsep{2pt}
\begin{eqnarray}
x_{D} & \approx & \sqrt{\frac{\hbar}{2\mu \Omega}}(f_{00}+f_{10}\hat{A}^{\dagger}+\hat{A}f_{01}+f_{20}\hat{A}^{\dagger 2}+\hat{A}^{2}f_{02}), \label{eq:xdmor} \\
p_{D} & \approx & i\sqrt{\frac{\hbar \mu \Omega}{2}}(g_{01}\hat{A}^{\dagger}+\hat{A}g_{01}+g_{20}\hat{A}^{\dagger 2}+\hat{A}^{2}g_{02}). \label{eq:pdmor}
\end{eqnarray}}
Here, the expansion coefficients $f_{ij}$ and $g_{ij}$ are functions of the excitation number operator $\hat{n}$ given by
{\setlength\arraycolsep{2pt}
\begin{eqnarray}
f_{00}(\hat{n}) & = & \sqrt{k}\left[f_{0}+\ln \left(\frac{(k-2)(k-\hat{n}-1)}{(k-1-2\hat{n})(k-2\hat{n})}\right)(1-\delta_{\hat{n},0})\right],\\
f_{10}(\hat{n}) & = & f_{01}(\hat{n})=\sqrt{\frac{k-1}{k}}\left(1+\frac{\hat{n}}{k-\hat{n}}\right),\\
f_{20}(\hat{n}) & = & f_{02}(\hat{n}) =\frac{k-1}{2 k \sqrt{k}}\left(\frac{-1}{(1-(\hat{n}-1)/k)(1-\hat{n}/k)} \right), \\
g_{10}(\hat{n}) & = & -g_{01}(\hat{n})=\sqrt{\frac{k-1}{k}}\left(\frac{k-2\hat{n}}{k-\hat{n}}\right), \\
g_{20}(\hat{n}) & = & -g_{02}(\hat{n})= -\frac{k-1}{k \sqrt{k}}\left(\frac{k-(2 \hat{n}-1)}{k(1-(\hat{n}-1)/k)(1-\hat{n}/k)} \right),
\end{eqnarray}}
where, in turn, 
\begin{equation}
f_{0}=\ln \ k-\left(\sum_{p=1}^{k-2}\frac{1}{p}-\Gamma \right),
\end{equation}
with
\begin{equation}
\Gamma=\lim_{m\to \infty}\left(\sum_{p=1}^{m}\frac{1}{p}-\ln \ m \right)=0.577216
\end{equation}
being the Euler constant, and $k=2 N+1=1/\chi_{a}$.

So then, the pair of approximate operators (\ref{eq:xdmor}) and (\ref{eq:pdmor}) are referred to as the deformed representation of the position and momentum. For the sake of simplicity, in subsequent calculations the natural units $\hbar=\mu$ ($\mu$ being the reduced mass of the molecule), together with $\Omega=1$, will be employed in (\ref{eq:xdmor}) and (\ref{eq:pdmor}). \\

The temporal evolution of our Morse-like coherent sates is obtained by application of the time evolution operator $\hat{U}(t)=e^{-i\hat{H}_{D}t/\hbar}$ upon the corresponding coherent state at $t=0$, where the deformed Hamiltonian $\hat{H}_{D}$ is given by Eq. (\ref{eq:spec1}). So, taking the LOQCS and DPSCS defined in (\ref{eq:edos}) and (\ref{eq:defps2}), respectively, as the initial states $|\alpha, \gamma; t=0 \rangle$, $|\zeta(\alpha), m; t=0 \rangle$, we get at time $t$ 
\begin{equation}\fl 
|\alpha,\gamma; t\rangle  =  \hat{U}(t)|\alpha,\gamma;0 \rangle  = e^{-i \Omega t(\hat{n}+1/2-(\hat{n}+1/2)^{2}\chi_{a}-\chi_{a}/4)}|\alpha, \gamma; 0 \rangle,
\end{equation}
\begin{equation}\fl 
|\zeta(\alpha),m; t\rangle  =  \hat{U}(t)|\zeta(\alpha), m; 0 \rangle  = e^{-i \Omega t(\hat{n}+1/2-(\hat{n}+1/2)^{2}\chi_{a}-\chi_{a}/4)}|\zeta(\alpha), m; 0 \rangle.
\end{equation}
Utilizing these states, together with the help of the expressions of the deformed observables $x_{D}$ and $p_{D}$ introduced above, we are now able to compute the average values of useful quantities including products of the type $x_{D}^{2}$, $p_{D}^{2}$, $x_{D}p_{D}$, and $p_{D}x_{D}$. That is, in addition to the average coordinate ($\langle x_{D} \rangle $) and momentum ($\langle p_{D} \rangle$) values, we can also calculate the  dispersions relations 
\begin{eqnarray}
\langle (\Delta x_{D})^{2} \rangle & = & \langle x_{D}^{2} \rangle-\langle x_{D} \rangle ^{2},\\
\langle (\Delta p_{D})^{2} \rangle & = & \langle p_{D}^{2} \rangle-\langle p_{D} \rangle ^{2}, \label{eq:disprel}
\end{eqnarray}
and, based on Eq. (\ref{eq:unrel}), examine the normalized uncertainty product
\begin{equation}
\Delta_{xp}=\frac{4 \langle (\Delta x_{D})^{2} \rangle \langle (\Delta p_{D})^{2} \rangle} {|\langle [x_{D},p_{D}]\rangle|^{2}}.
\label{eq:nup}
\end{equation}
On the basis of these relations, any one of the deformed coherent states defined here in which $\langle (\Delta O)^{2} \rangle <0.5$, where $O$ is either $x_{D}$ or $p_{D}$, will be called a squeezed state. On the other hand, the state will be called a minimum-uncertainty coherent state if the equality $\Delta_{xp}=1$ holds. \\

\begin{figure}[h!]
\begin{center}
\includegraphics[width=11cm, height=13cm]{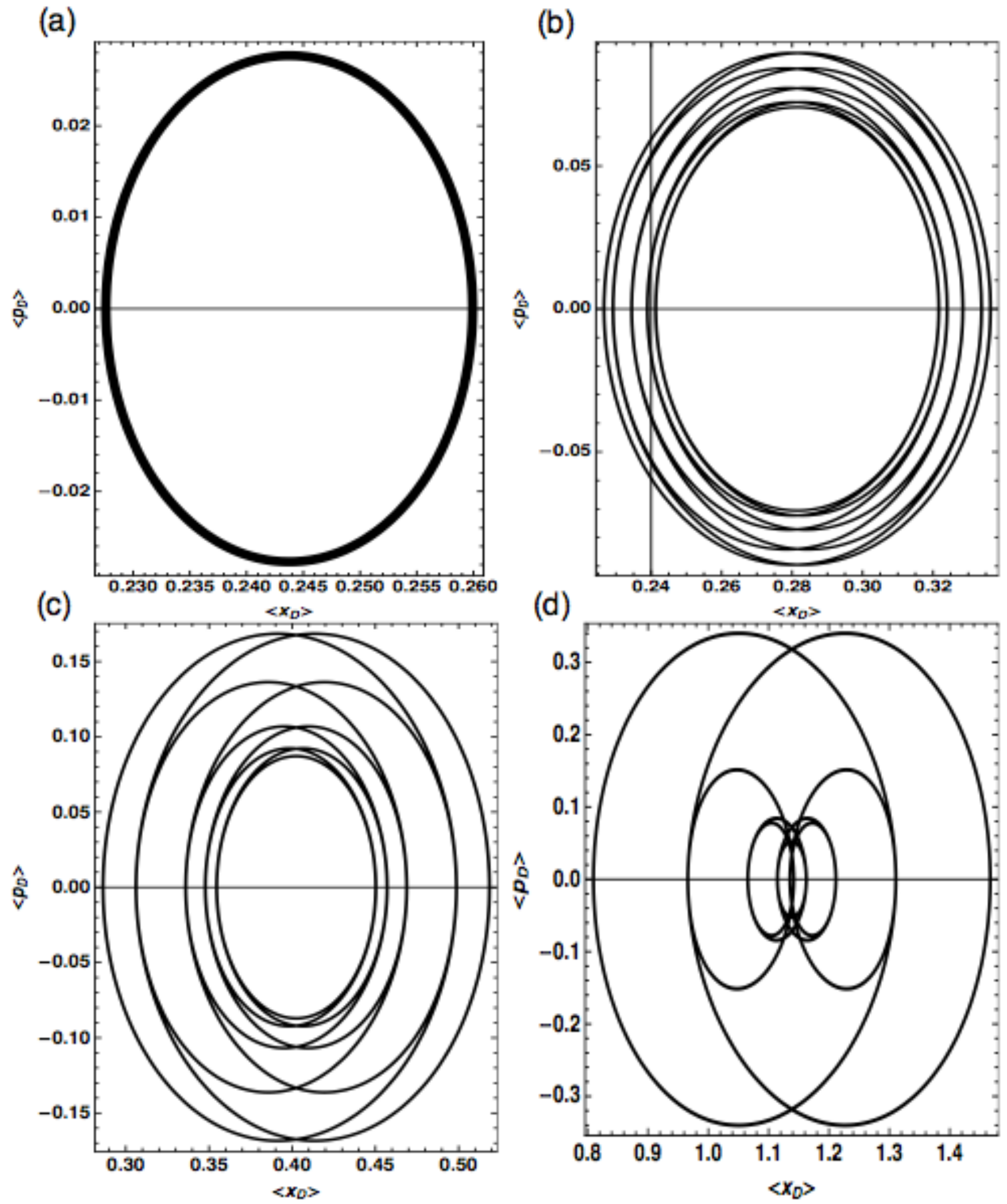} 
\caption{Phase space trajectories of the Morse-like squeezed vacuum state $|\alpha=0,\gamma \rangle$. Frames (a) to (d) correspond to $\gamma= 0.1$, $0.3$, $0.5$ and $0.8$, respectively.}
\label{fig:phasespace1}
\end{center}
\end{figure}

\begin{figure}[h!]
\begin{center}
\includegraphics[width=14cm, height=10cm]{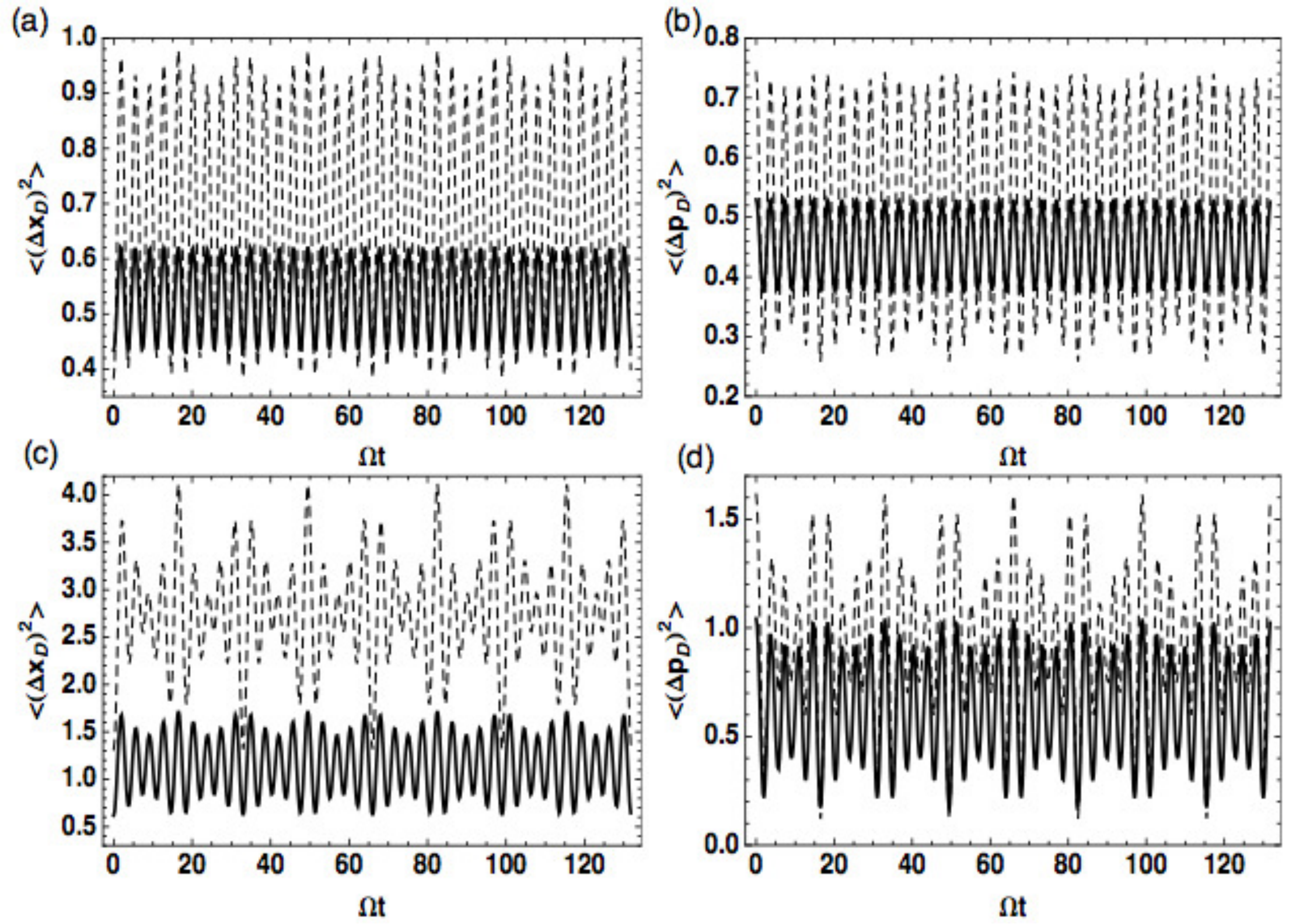} 
\caption{Temporal evolution of the uncertainties in coordinate (frames (a), (c)) and momentum (frames (b), (d)) for the Morse-like squeezed vacuum state $|\alpha=0,\gamma \rangle$. Frames (a) and (b) correspond to $\gamma=0.1$ (continuous line) and $\gamma=0.3$ (dashed line). Frames (c) and (d) correspond to $\gamma=0.5$ (continuous line) and $\gamma=0.8$ (dashed line).}
\label{fig:disprelationsgammat}
\end{center}
\end{figure}

\begin{figure}[h!]
\begin{center}
\includegraphics[width=14cm, height=5cm]{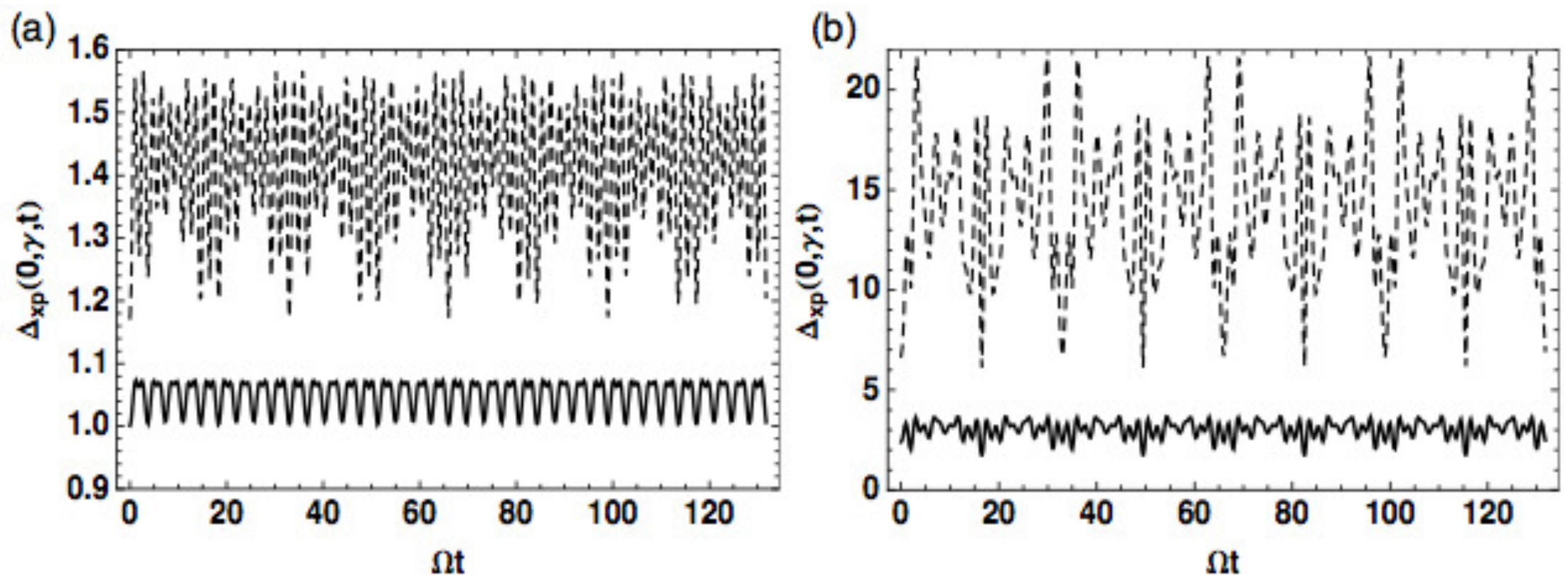} 
\caption{Temporal evolution of the normalized uncertainty product $\Delta_{xp}=4 \langle (\Delta x_{D})^{2} \rangle \langle (\Delta p_{D})^{2} \rangle /|\langle [x_{D},p_{D}]\rangle|^{2}$ for the Morse-like squeezed vacuum state $|\alpha=0,\gamma \rangle$. Frame (a) correspond to $\gamma=0.1$ (continuous line) and $\gamma=0.3$ (dashed line). Frame (b) correspond to $\gamma=0.5$ (continuous line) and $\gamma=0.8$ (dashed line).}
\label{fig:ndisprelationsgammat}
\end{center}
\end{figure}

Phase space trajectories (computed by plotting $\langle x_{D} \rangle $ against $\langle p_{D} \rangle$) for the LOQCS with $\alpha=0$ (the Morse-like vacuum state $|\alpha=0,\gamma\rangle$) are shown in Fig. \ref{fig:phasespace1} for the values of the squeezing parameter quoted in Fig. \ref{fig:statistics2} ($\gamma=0.1$, $0.3$, $0.5$, and $0.8$). It is found that if the squeezing parameter is sufficiently small, say $\gamma = 0.1$, see Fig. \ref{fig:phasespace1}(a), the evolution of the state appears to be well localized on phase space. However, as the value of $\gamma$ increases (see for instance Fig. \ref{fig:phasespace1}(b) to (c)) the state exhibits a less stable behavior. Then, for $\gamma =0.8$ (see Fig. \ref{fig:phasespace1}(d)) its temporal stability is altogether lost. Of course, this last fact is closely related to spreading effects of the associated coherent state wave packet, which is manifested in Figs. \ref{fig:disprelationsgammat} and \ref{fig:ndisprelationsgammat} where we display the corresponding temporal behavior of the variances (\ref{eq:disprel}), and the normalized uncertainty product (\ref{eq:nup}), respectively. Following the sequence of graphs in Fig. \ref{fig:disprelationsgammat}, we can see that the larger the value of the squeeze parameter $\gamma$ is, the larger the oscillatory amplitude of both fluctuations $\langle (\Delta x_{D})^{2} \rangle$ and $\langle (\Delta p_{D})^{2} \rangle$ is. The increase in $\gamma$ gives rise to a temporal behavior that seems like a sequence of beats (by beats, we refer to the times at which the amplitude of fluctuations is more pronounced). These beats take place simultaneously with the presence of the inner trajectories on phase space seen in Fig \ref{fig:phasespace1}. Besides, we observe that the squeezing of both $x_{D}$ and $p_{D}$ is more pronounced at certain stages of state's evolution for sufficiently small values of $\gamma$ (see Fig. \ref{fig:disprelationsgammat}(a) and (b)). On the other hand, with the increase in $\gamma$ the state cease to be squeezed with respect to $x_{D}$ (Fig. \ref{fig:disprelationsgammat}(c)), but it is still squeezed on the momentum $p_{D}$ (see Fig. \ref{fig:disprelationsgammat}(d)). From the lower graph of Fig. \ref{fig:ndisprelationsgammat}(a) we can observe that although the squeezed vacuum state is almost a minimum-uncertainty state at certain time instants (i.e., $\Delta_{xp}\approx 1$) for $\gamma= 0.1$, this is not so for larger values of $\gamma$. 

\begin{figure}[h!]
\begin{center}
\includegraphics[width=15cm, height=4cm]{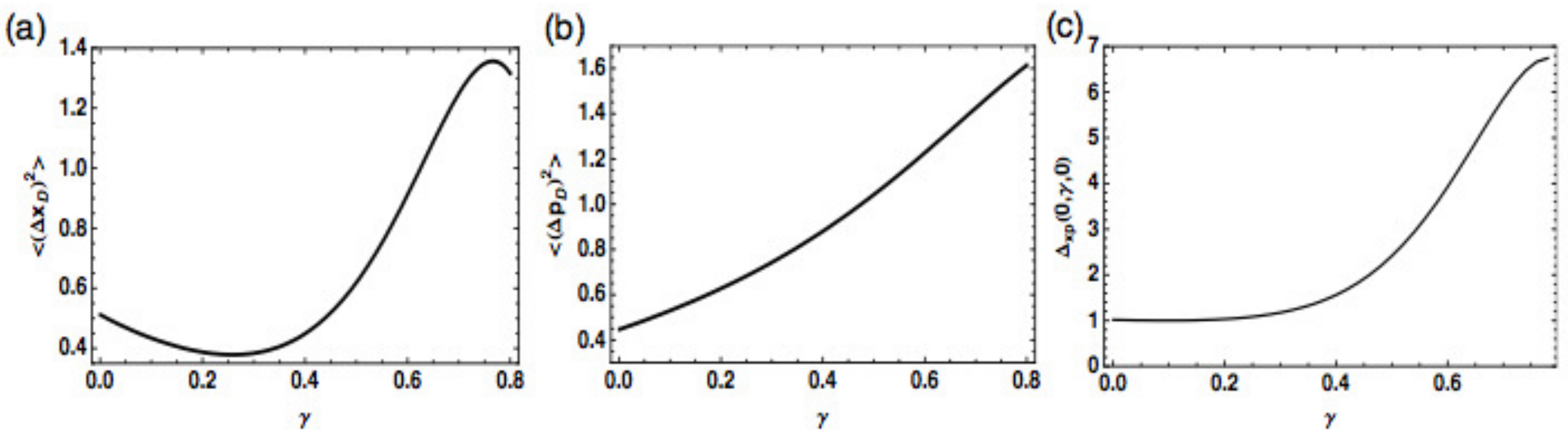} 
\caption{Uncertainty in coordinate $\langle (\Delta x_{D})^{2} \rangle$ (a), in momentum $\langle (\Delta p_{D})^{2} \rangle$ (b), and the corresponding normalized uncertainty product $\Delta_{xp}=4 \langle (\Delta x_{D})^{2} \rangle \langle (\Delta p_{D})^{2} \rangle /|\langle [x_{D},p_{D}]\rangle|^{2}$ (c) at time $t=0$ as a function of $\gamma$ for the Morse-like squeezed vacuum state $|\alpha=0,\gamma \rangle$.}
\label{fig:disprelationsgamma}
\end{center}
\end{figure}

In Fig. \ref{fig:disprelationsgamma} we have plotted the behavior of the dispersions $\langle (\Delta x_{D})^{2} \rangle$ and $\langle (\Delta p_{D})^{2} \rangle$ and the normalized uncertainty product $\Delta_{xp}$ at $t=0$ as functions of $\gamma$ for the vacuum state $|\alpha=0,\gamma \rangle$. We see that for $\gamma < 0.4$ the state can show squeezing in $x_{D}$ (Fig. \ref{fig:disprelationsgamma}(a)). With respect to $p_{D}$, the squeezing is restricted to a much smaller region  of $\gamma$. And, at least for $0 < \gamma < 0.2$, where the uncertainty product $\Delta_{xp}$ is almost $1$ (see Fig. \ref{fig:disprelationsgamma}(c)) the state may be regarded as a minimum-uncertainty coherent state. \\

\begin{figure}[h!]
\begin{center}
\includegraphics[width=10cm, height=8.5cm]{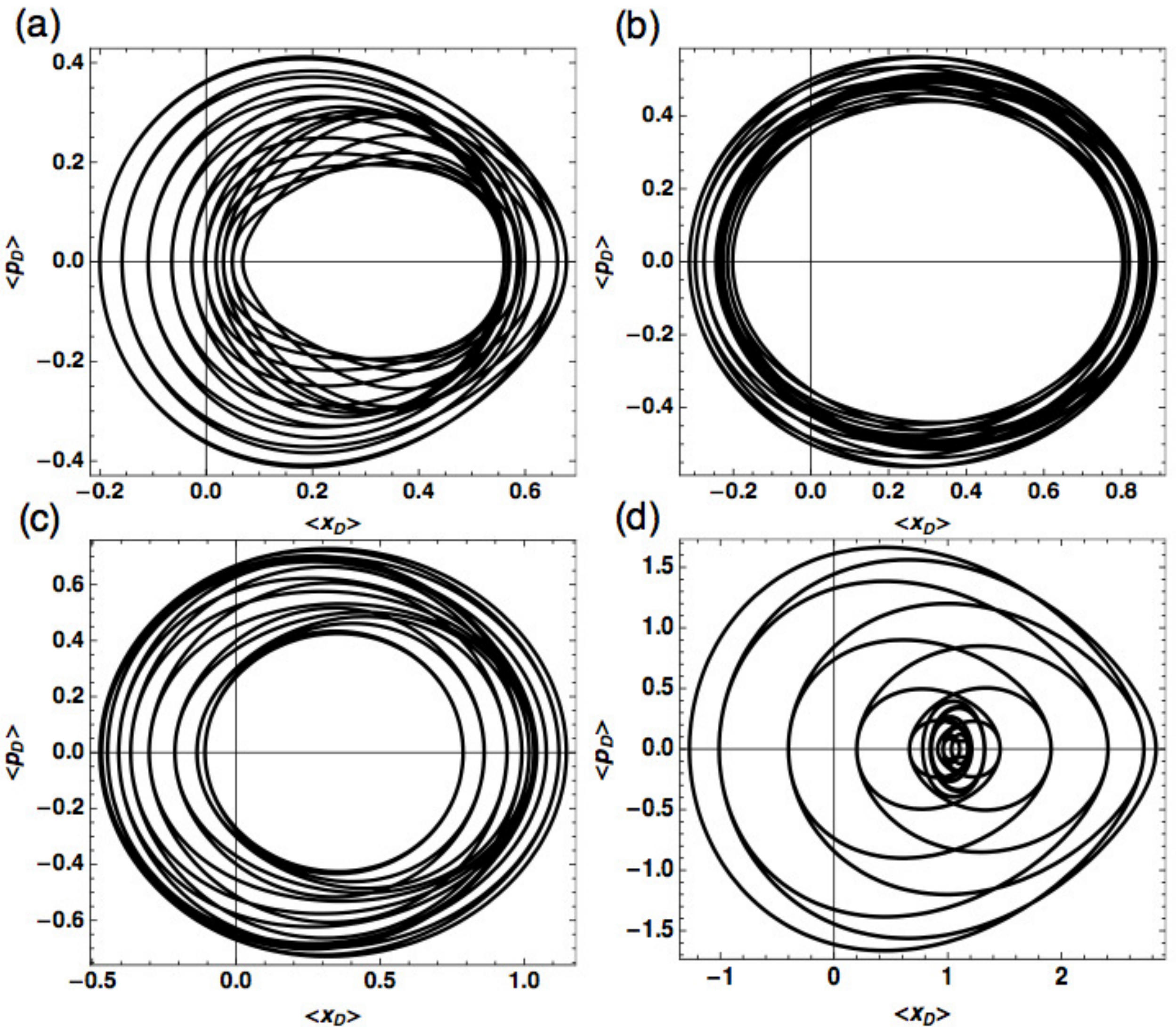} 
\caption{Phase space trajectories for Morse-like squeezed coherent states, $|\alpha,\gamma \rangle$, with $\gamma=0.3$. Frames (a) to (d) correspond to $\alpha= 0.2$, $0.4$, $0.8$ and $1.6$, respectively.}
\label{fig:phasespace2}
\end{center}
\end{figure}

\begin{figure}[h!]
\begin{center}
\includegraphics[width=14cm, height=10cm]{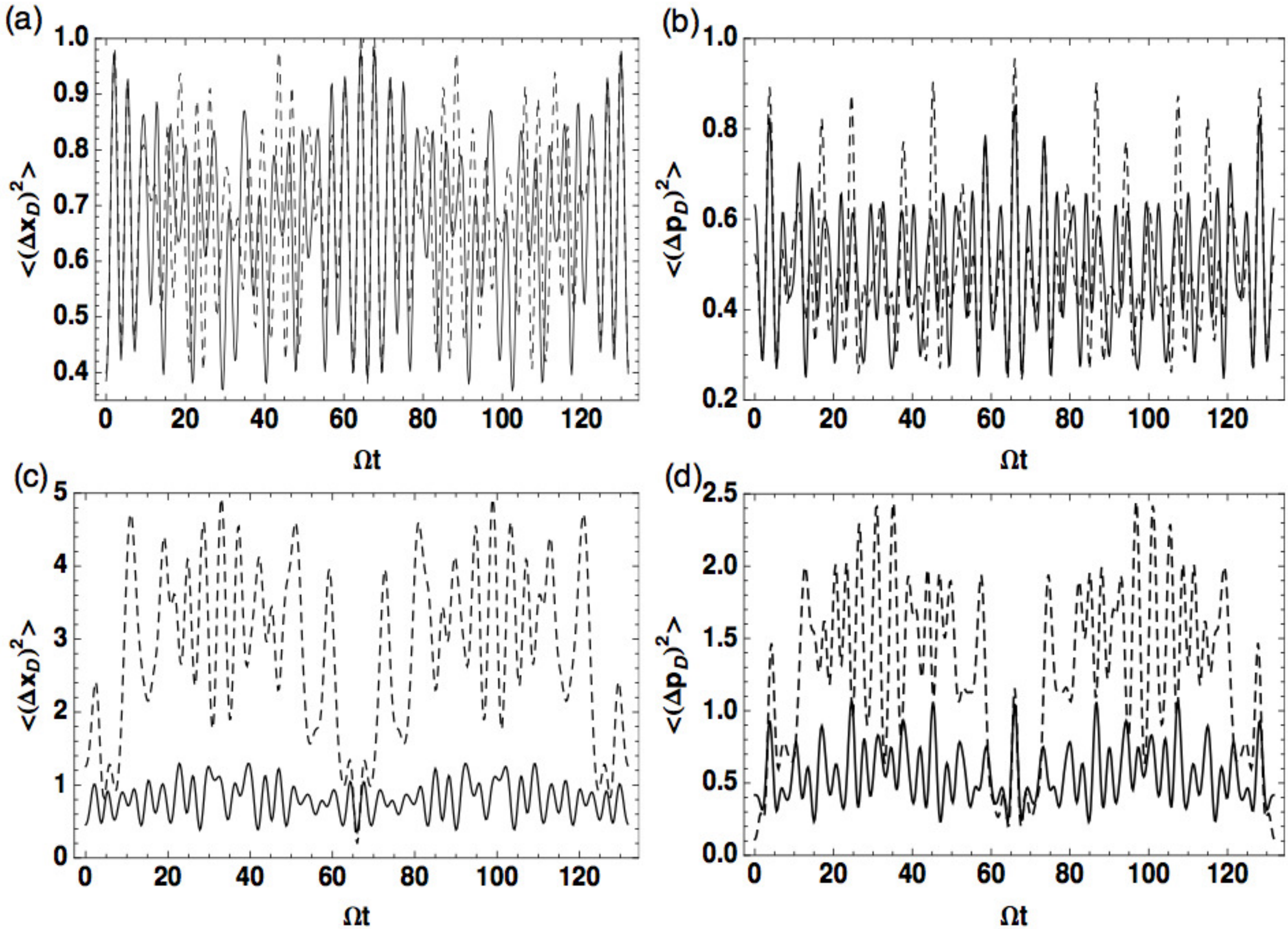} 
\caption{Temporal evolution of the uncertainties in coordinate (frames (a), (c)) and momentum (frames (b), (d)) for the LOQCS $|\alpha,\gamma=0.3 \rangle$. Frames (a) and (b) correspond to $\alpha=0.2$ (continuous line) and $\alpha=0.4$ (dashed line). Frames (c) and (d) correspond to $\alpha=0.8$ (continuous line) and $\alpha=1.6$ (dashed line).}
\label{fig:disprelationsalphat}
\end{center}
\end{figure}

\begin{figure}[h!]
\begin{center}
\includegraphics[width=14cm, height=5cm]{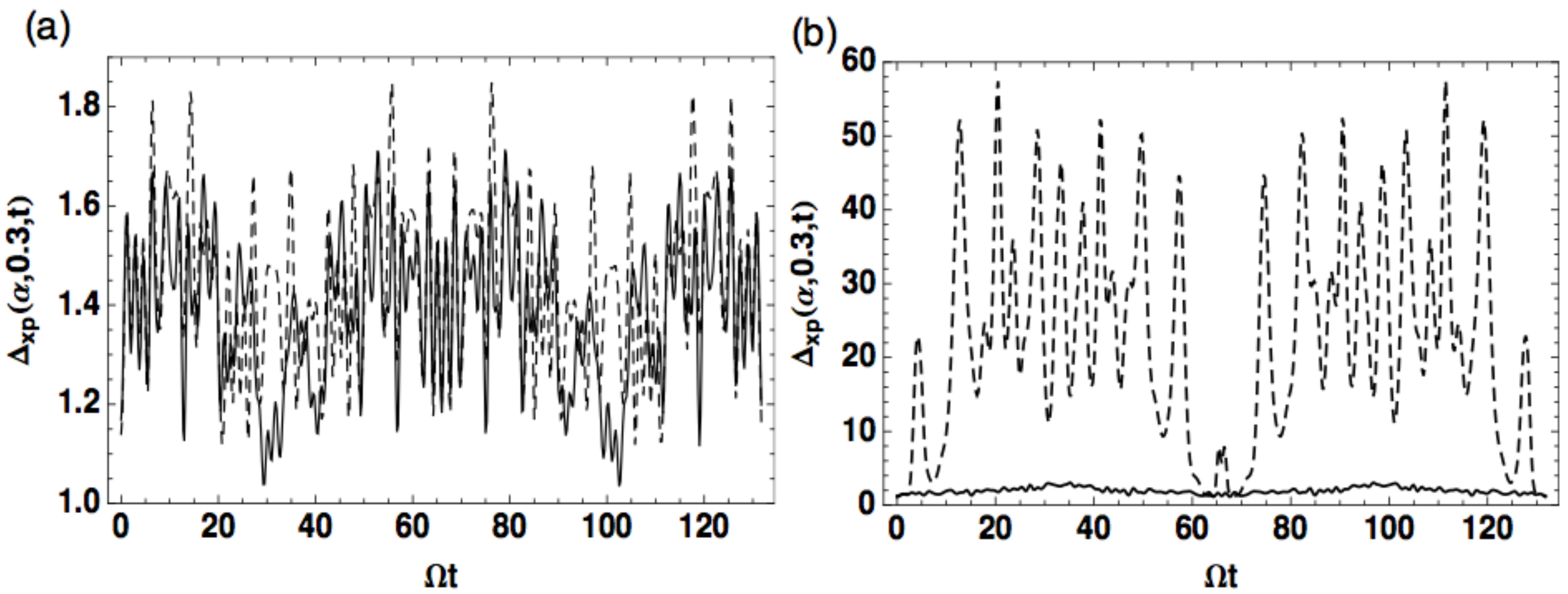} 
\caption{Temporal evolution of the normalized uncertainty product $\Delta_{xp}=4 \langle (\Delta x_{D})^{2} \rangle \langle (\Delta p_{D})^{2} \rangle /|\langle [x_{D},p_{D}]\rangle|^{2}$ for the LOQCS  $|\alpha,\gamma=0.3 \rangle$. Frame (a) correspond to $\alpha=0.2$ (continuous line) and $\alpha=0.4$ (dashed line). Frame (b) correspond to $\alpha=0.8$ (continuous line) and $\alpha=1.6$ (dashed line).}
\label{fig:ndisprelationsalphat}
\end{center}
\end{figure}

Let us regard the more general case of the LOQCS for different values of $\alpha$. In Fig. \ref{fig:phasespace2}, for example, unlike the case $\alpha=0$ described above, one sees that the corresponding trajectories on phase space turn out to have an ovoid shape along the horizontal axis associated with the expectation value of $x_{D}$. We see from Figures \ref{fig:phasespace2} (a)- \ref{fig:phasespace2}(b), for $\alpha=0.2$ and $0.4$, that the states appear to be approximately well-localized in time, whereas for larger values, starting around $0.8$, the loss of localization becomes more pronounced. This fact is understandable form the point of view of the corresponding dispersion relations and the uncertainty product displayed in Figs.  \ref{fig:disprelationsalphat} and \ref{fig:ndisprelationsalphat}, respectively, for the said values of $\alpha$. Notice that the larger the value of $\alpha$ is, the larger the amplitude of the fluctuations is; this is particularly highlighted by going from $\alpha=0.8$ to $1.6$ (see Fig. \ref{fig:disprelationsalphat}(c) and (d)). In addition, the LOQCS can display squeezing in both $x_{D}$ and $p_{D}$, which is restricted to certain periods of time. Nevertheless, note that the larger the value of $\alpha$, the smaller the time intervals at which the squeezing is present. It is seen from Fig. \ref{fig:ndisprelationsalphat} that most of the time the LOQCS are not minimum-uncertainty coherent states. The large increment of the uncertainty product exhibited in Fig. \ref{fig:ndisprelationsalphat}(b) explains why the inner trajectories in the phase space seen in Fig. \ref{fig:phasespace2}(d) collapse faster. \\

\newpage

\begin{figure}[h!]
\begin{center}
\includegraphics[width=10cm, height=9cm]{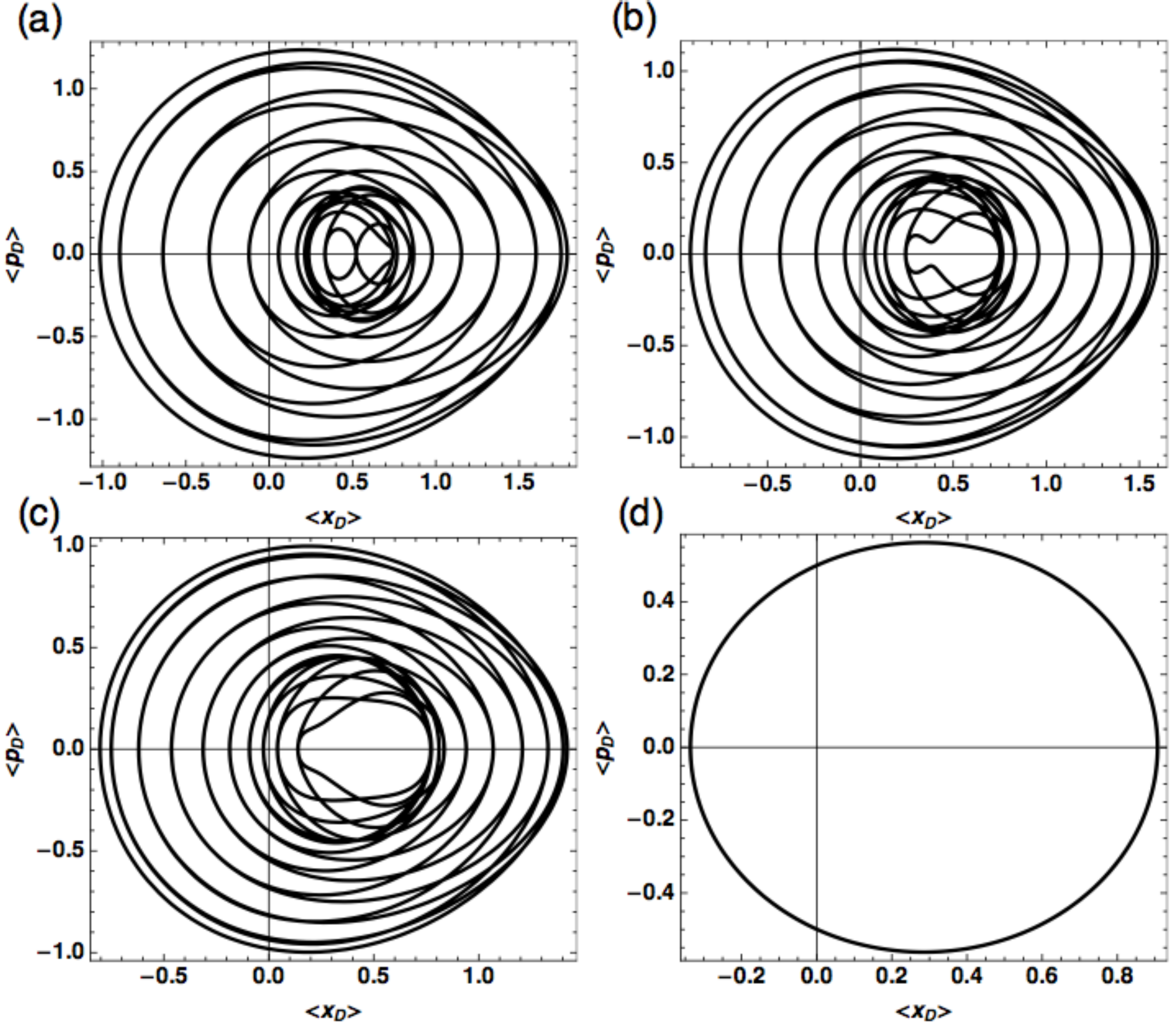} 
\caption{Phase space trajectories for deformed photon-subtracted coherent states, $|\zeta(\alpha),m\rangle$, with $\alpha=1$. Frames (a) to (d) correspond to $m= 0$, $2$, $4$ and $8$, respectively.}
\label{fig:phasespace3}
\end{center}
\end{figure}

\begin{figure}[h!]
\begin{center}
\includegraphics[width=14cm, height=10cm]{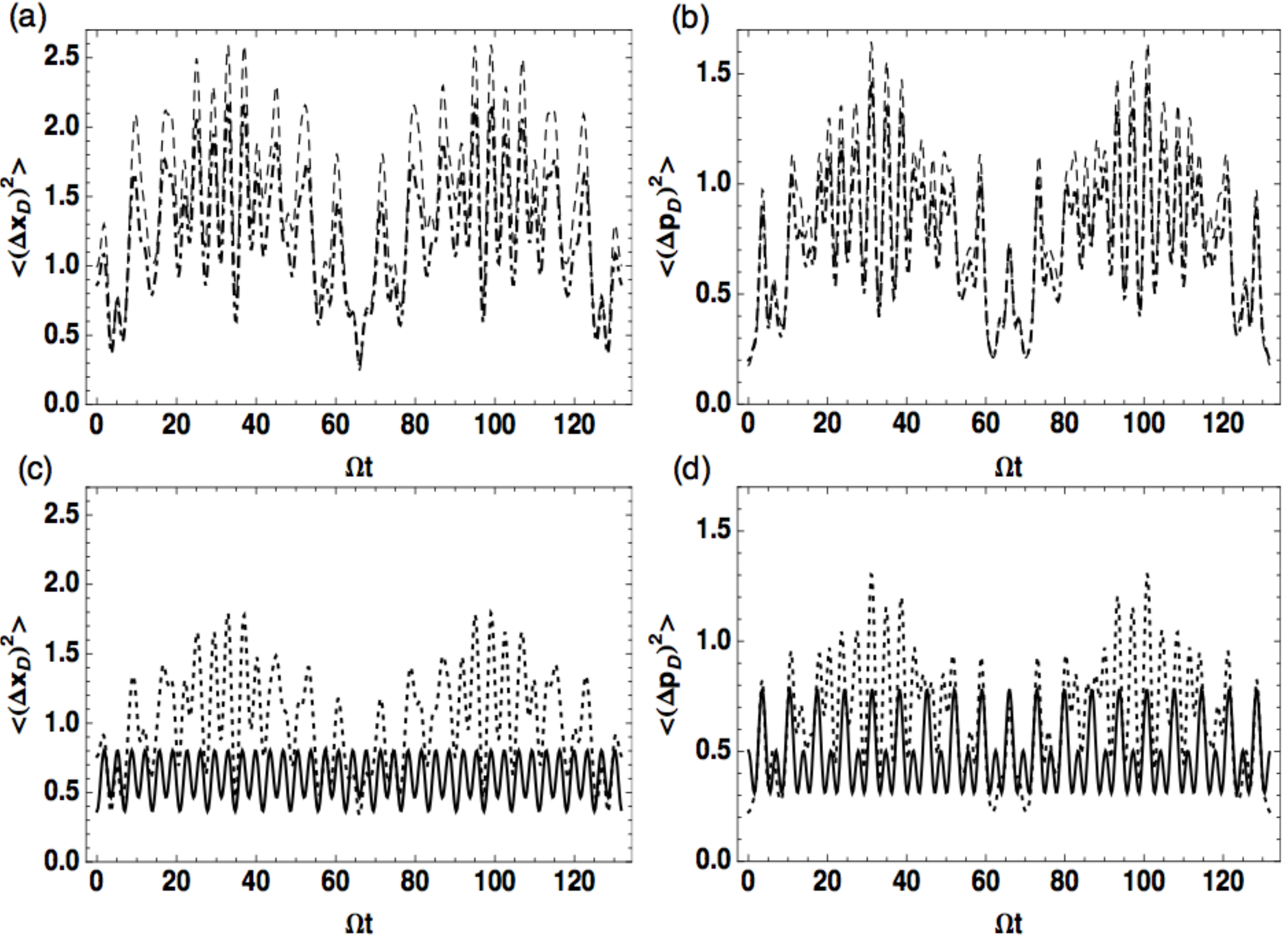} 
\caption{Temporal evolution of the uncertainties in coordinate (frames (a), (c)) and momentum (frames (b), (d)) for a deformed photon-subtracted coheret state $|\zeta(\alpha),m \rangle$ with $\alpha=1$. Frames (a) and (b) correspond to $m=0$ (dashed line) and $m=2$ (dot-dashed line). Frames (c) and (d) correspond to $m=4$ (dotted line) and $m=8$ (continuous line).}
\label{fig:disprelationsmt}
\end{center}
\end{figure}

\begin{figure}[h!]
\begin{center}
\includegraphics[width=14cm, height=5cm]{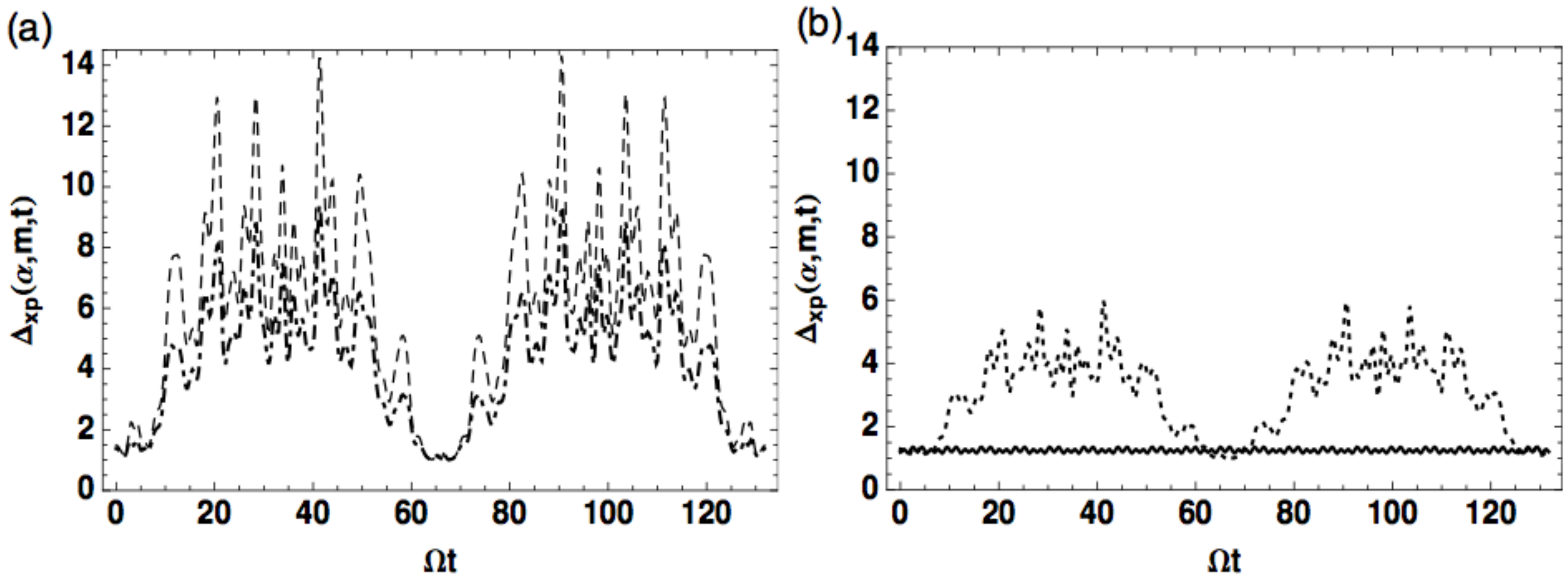} 
\caption{Temporal evolution of the normalized uncertainty product $\Delta_{xp}=4 \langle (\Delta x_{D})^{2} \rangle \langle (\Delta p_{D})^{2} \rangle /|\langle [x_{D},p_{D}]\rangle|^{2}$ for a deformed  photon-subtracted coherent state $|\zeta (\alpha),m \rangle$ with $\alpha=1$. Frame (a) correspond to $m=0$ (dashed line) and $m=2$ (dot-dashed line). Frame (b) correspond to $m=4$ (dotted line) and $m=8$ (continuous line).}
\label{fig:ndisprelationsmt}
\end{center}
\end{figure}

Let us now examine the temporal features of the deformed photon-subtracted coherent states (DPSCSs) defined by Eq. (\ref{eq:defps2}). Figure \ref{fig:phasespace3} illustrates the phase space trajectories of such states for the values of the parameter $m$ (the number of depleted quanta) quoted in Fig. \ref{fig:statistics6} ($m=0$, $2$, $4$, and $8$, for fixed $\alpha=1$). It is found that the trajectories of the states maintain their characteristic ovoid shape for each value of $m$. In addition, as the number of depleted quanta increases, it appears that the DPSCSs tend to be well localized with time (this tendency was confirmed for several values of $\alpha$), which is in accordance with the results for the dispersion relations displayed in Figs. \ref{fig:disprelationsmt}(a)-(d). Effectively, from the graphs in Figs. \ref{fig:disprelationsmt}(c) and (d) we can see that for high values of the parameter $m$, starting around $4$ depleted quanta, the dispersion of $x_{D}$ and $p_{D}$ shows a reduced oscillatory amplitude. On the other hand, note also that within  certain time intervals, such states can display squeezing both in the position and momentum variables, being a little more noticeable for the latter (see Fig. \ref{fig:disprelationsmt}(d)). The decreasing oscillatory amplitude of the normalized uncertainty product potted in Fig. \ref{fig:ndisprelationsmt} reveals the fact that the larger the value of $m$ is, the more localized in time is the corresponding photon-subtracted coherent state. In particular, for $m=8$, see Fig. \ref{fig:ndisprelationsmt}(b), we can observe that the uncertainty product takes values very close to unity, in which case the state is `almost' a minimum-uncertainty coherent state. This fact is consistent with state's phase space behavior seen in Fig. \ref{fig:phasespace3}(d). Again, those times at  which the amplitude of the uncertainty product takes its largest values turn out to be the periods over which the phase space trajectories reach very small amplitudes, as seen in Figs. \ref{fig:phasespace3}(a) and (b).

\begin{figure}[h!]
\begin{center}
\includegraphics[width=14cm, height=15cm]{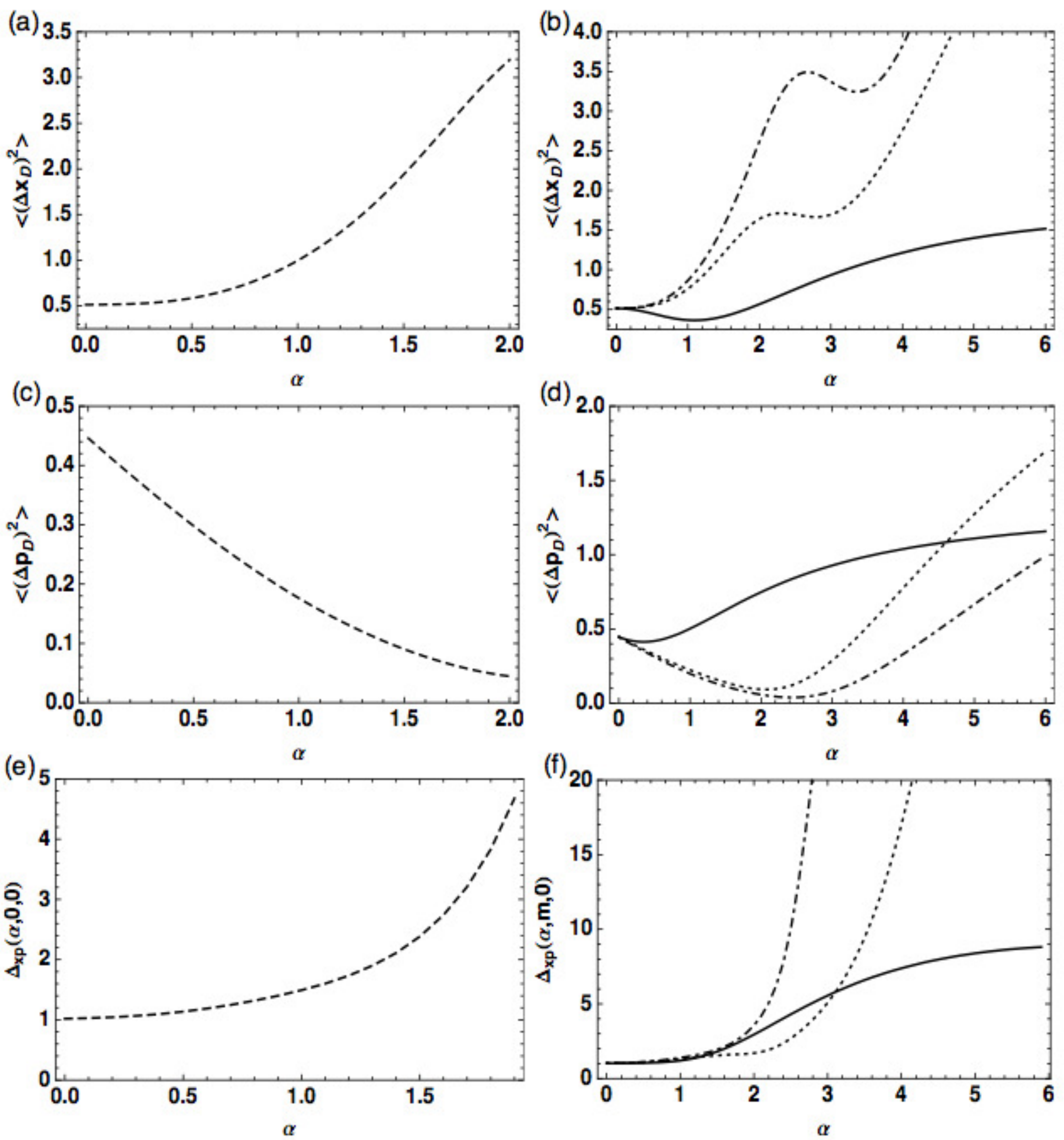} 
\caption{Dispersion in coordinate $\langle (\Delta x_{D})^{2} \rangle$ (frames (a)-(b)), in momentum $\langle (\Delta p_{D})^{2} \rangle$ (frames (c)-(d)), and the corresponding normalized uncertainty product $\Delta_{xp}=4 \langle (\Delta x_{D})^{2} \rangle \langle (\Delta p_{D})^{2} \rangle /|\langle [x_{D},p_{D}]\rangle|^{2}$ (frames (e)-(f)) at time $t=0$ as a function of $\alpha$ for a deformed photon-subtracted coherent state $|\zeta(\alpha),m \rangle$ with $m=0$ (dashed line), $m=2$ (dot-dashed line), $m=4$ (dotted line), and $m=8$ (continuous line). }
\label{fig:disprelationsmalpha}
\end{center}
\end{figure}

In Fig. \ref{fig:disprelationsmalpha} we show the dispersions and uncertainty relations of the DPSCSs at $t=0$ as functions of $\alpha$ for various values of $m$. For $m=0$, when the DPSCSs reduce to the DOCSs, the behavior of the dispersion in $x_{D}$ is similar to the harmonic case only for $\alpha <0.5$ (Fig. \ref{fig:disprelationsmalpha}(a)), whereas for the case of $p_{D}$ the dispersion takes values smaller than $0.5$ in a wider interval of $\alpha$, indicating the presence of squeezing (Fig. \ref{fig:disprelationsmalpha}(c)). For $0<\alpha<0.5$, approximately, the states are almost minimum-uncertainty states, as seen in Fig. \ref{fig:disprelationsmalpha}(e). Curiously, it was found that just for higher values of $m$, say $m=8$ (see Fig. \ref{fig:disprelationsmalpha}(b)), there exists a small region of $\alpha$ where the squeezing in $x_{D}$ is present; for intermediate values of $m$ the dispersion in $x_{D}$ increases drastically as $\alpha$ increases. On the other hand, with respect to the dispersion in $p_{D}$, the states exhibit significant squeezing for intermediate values of $m$, whereas for higher values the squeezing is almost lost (see Fig. \ref{fig:disprelationsmalpha}(d)). From the normalized uncertainty product displayed in Fig. \ref{fig:disprelationsmalpha}(f) we see that the DPSCSs at $t=0$ are minimum-uncertainty coherent states at least for $0<\alpha<1$. For the intermediate values $m=2$ and $4$, the uncertainty product increases very quickly for $\alpha >1$, but for higher values of $m$, say $m=8$, it increases slowly and displays a significant reduction. This last fact explains why the states thus constructed become more localized for increasing values of $m$.

\section{Conclusions}

Based on the f-deformed oscillator formalism, we have introduced two types of squeezed coherent states associated with a Morse system (Morse-like squeezed coherent states) and have examined them from a statistical and dynamical viewpoint. Such states were constructed considering the two following definitions: i) as quasi-eigenstates of f-deformed ladder operators, called deformed ladder-operator quasi-coherent states (LOQCSs), and ii) as deformed photon-subtracted coherent states (DPSCSs); the latter being constructed by applying iteratively the deformed annihilation operator $\hat{A}=\hat{a}f(\hat{n})$ upon an initial state that is considered to be a deformed displacement operator coherent state (DOCS). For both types of squeezed coherent states, we have illustrated their statistical properties and have evaluated the temporal evolution of some quantities of physical interest like phase space trajectories, dispersion relations, and squeezing. In short, we emphasize the main features of the introduced states:

\begin{itemize}
\item Concerning their statistical properties, we have found that, depending on the values of the parameters $\alpha$ and $\gamma$, the LOQCSs $|\alpha,\gamma\rangle$ can exhibit both sub-Poissonian and super-Poissonian statistics. In particular, for $\alpha=0$, the Morse-like vacuum state $|0,\gamma \rangle$ always exhibits super-Poissonian statistics as a function of $\gamma$, which leads to a faster spreading of state's occupation number distribution function. On the other hand, it was found that the DPSCSs $|\zeta(\alpha),m \rangle$ exhibit sub-Poissonian statistics as a function of $\alpha$ for all $m$.
\item Only for sufficiently small values of  $\alpha$ and $\gamma$, the evolution of the LOQCSs show a somewhat well-localized behavior on phase space. As we have seen, these states are almost minimum-uncertainty coherent states under these conditions. On the contrary, for higher values of $\alpha$ and/or $\gamma$ the dispersion of both the coordinate and momentum increases and accordingly the deformation and collapse of states' phase space trajectories become much more significant. This effect is due to the influence of the nonlinearity of the Hamiltonian model. The LOQCSs also display squeezing in the deformed coordinate and momentum variables at certain stages of their evolution and for certain restricted values of $\gamma$.
\item Finally, it is worth mentioning that the process of subtracting $m$ quanta from the DOCSs, from which the concept of DPSCSs arises, turn out to have noticeable effects on the reduction of the dispersion of both the coordinate and momentum. On the one hand, the states thus obtained tend to be well-localized on phase space for high values of $m$ and, on the other hand, they exhibit a significant increment of squeezing as a function of $\alpha$ for intermediate values of $m$. Indeed, the DPSCSs deserve to be classed as squeezed states.
\end{itemize}

\newpage

\ack This work was partially supported by DGAPA UNAM project No. IN 108413 and by CONACyT-M\'{e}xico through project 166961

\end{document}